\documentclass[article,longauth]{aa} 
\usepackage{graphicx}
\usepackage{xcolor,colortbl}
\usepackage[]{natbib}
\usepackage{multirow}
\usepackage{amsmath} 
\usepackage{txfonts}  
\usepackage{enumitem}
\usepackage{orcidlink}
\usepackage{multirow}

\def\degb{^{\circ}}

\begin{document}

\title{Imaging detection of the inner dust belt and the four exoplanets in the HR\,8799 system with JWST's MIRI coronagraph}

\author{
    Anthony Boccaletti\orcidlink{0000-0001-9353-2724}\inst{\ref{lesia}}, 
    Mathilde Mâlin\inst{\ref{lesia}}, 
    Pierre Baudoz\inst{\ref{lesia}}, 
    Pascal Tremblin\orcidlink{0000-0001-6172-3403}\inst{\ref{cea}}, 
    Clément Perrot\orcidlink{0000-0003-3831-0381}\inst{\ref{lesia}},
    Daniel Rouan\inst{\ref{lesia}}, 
    Pierre-Olivier Lagage\inst{\ref{cea}}, Niall Whiteford\inst{\ref{roe}},
    Paul Mollière\orcidlink{0000-0003-4096-7067}\inst{\ref{mpia}},
    Rens Waters\inst{\ref{radboud},\ref{hfml},\ref{sron}},
    Thomas Henning\inst{\ref{mpia}}, 
    Leen Decin\inst{\ref{leuven}},
    Manuel G\"udel\inst{\ref{vienna},\ref{mpia},\ref{eth}},
    Bart Vandenbussche\orcidlink{0000-0002-1368-3109}\inst{\ref{leuven}},
    Olivier Absil\inst{\ref{star}},
    Ioannis Argyriou\orcidlink{0000-0003-2820-1077}\inst{\ref{leuven}}, 
    Jeroen Bouwman\orcidlink{0000-0003-4757-2500}\inst{\ref{mpia}},
    Christophe Cossou\orcidlink{0000-0001-5350-4796}\inst{\ref{cea}},
    Alain Coulais\orcidlink{0000-0001-6492-7719}\inst{\ref{cea},\ref{lerma}}, 
    Ren\'e Gastaud\inst{\ref{parissaclay}},
    Alistair Glasse\orcidlink{0000-0002-2041-2462}\inst{\ref{ukatc}},
    Adrian M. Glauser\orcidlink{0000-0001-9250-1547}\inst{\ref{eth}},
    Inga Kamp\orcidlink{0000-0001-7455-5349}\inst{\ref{kapteyn}},
    Sarah Kendrew\orcidlink{0000-0002-7612-0469}\inst{\ref{esa}},
    Oliver Krause\inst{\ref{mpia}},
    Fred Lahuis\inst{\ref{sron}},
    Michael Mueller\orcidlink{0000-0003-3217-5385}\inst{\ref{kapteyn}},
    Goran Olofsson\orcidlink{0000-0003-3747-7120}\inst{\ref{stockholm}},
    Polychronis Patapis\orcidlink{0000-0001-8718-3732}\inst{\ref{eth}},
    John Pye\inst{\ref{leicester}},
    Pierre Royer\orcidlink{0000-0001-9341-2546}\inst{\ref{leuven}},
    Eugene Serabyn\inst{\ref{jpl}},
    Silvia Scheithauer\orcidlink{0000-0003-4559-0721}\inst{\ref{mpia}},
    Luis Colina\orcidlink{0000-0002-9090-4227}\inst{\ref{cab}},
    Ewine F. van Dishoeck\orcidlink{0000-0001-7591-1907}\inst{\ref{leiden}},
    G\"oran Ostlin\inst{\ref{oskar}},
    Tom P.\ Ray\inst{\ref{dublin}},
    Gillian Wright\inst{\ref{ukatc}}
}

\institute{
LESIA, Observatoire de Paris, Universit{\'e} PSL, CNRS, Sorbonne Universit{\'e}, Univ. Paris Diderot, Sorbonne Paris Cit{\'e}, 5 place Jules Janssen, 92195 Meudon, France\label{lesia}
\and Universit{\'e} Paris-Saclay, Universit{\'e} Paris Cit{\'e}, CEA, CNRS, AIM, 91191, Gif-sur-Yvette, France\label{cea}
\and  Max-Planck-Institut f\"ur Astronomie (MPIA), K\"onigstuhl 17, 69117 Heidelberg, Germany \label{mpia}
\and  Institute for Astronomy, University of Edinburgh, Royal Observatory, Blackford Hill, Edinburgh EH9 3HJ\label{roe}
\and  Department of Astrophysics/IMAPP, Radboud University, PO Box 9010, 6500 GL Nijmegen, the Netherlands\label{radboud}
\and  HFML - FELIX. Radboud University PO box 9010, 6500 GL Nijmegen, the Netherlands\label{hfml}
\and  SRON Netherlands Institute for Space Research, Niels Bohrweg 4, 2333 CA Leiden, the Netherlands\label{sron}
\and   Department of Astrophysics, University of Vienna, T\"urkenschanzstrasse 17, 1180 Vienna, Austria\label{vienna}
\and ETH Z\"urich, Institute for Particle Physics and Astrophysics, Wolfgang-Pauli-Strasse 27, 8093 Z\"urich, Switzerland\label{eth}
\and Institute of Astronomy, KU Leuven, Celestijnenlaan 200D, 3001 Leuven, Belgium\label{leuven}
\and STAR Institute, Universit\'e de Li\`ege, All\'ee du Six Ao\^ut 19c, 4000 Li\`ege, Belgium\label{star}
\and Universit\'e Paris-Saclay, CEA, D\'epartement d'Electronique des D\'etecteurs et d'Informatique pour la Physique, 91191, Gif-sur-Yvette, France\label{parissaclay}
\and LERMA, Observatoire de Paris, Universit\'e PSL, Sorbonne Universit\'e, CNRS, Paris, France\label{lerma}
\and UK Astronomy Technology Centre, Royal Observatory, Blackford Hill, Edinburgh EH9 3HJ, UK\label{ukatc}
\and  Kapteyn Institute of Astronomy, University of Groningen, Landleven 12, 9747 AD Groningen, the Netherlands\label{kapteyn}
\and European Space Agency, Space Telescope Science Institute, Baltimore, MD, USA\label{esa}
\and Department of Astronomy, Stockholm University, AlbaNova University Center, 10691 Stockholm, Sweden\label{stockholm}
\and  School of Physics \& Astronomy, Space Research Centre, Space Park Leicester, University of Leicester, 92 Corporation Road, Leicester, LE4 5SP, UK\label{leicester}
\and  Centro de Astrobiología (CAB), CSIC-INTA, ESAC Campus, Camino Bajo del Castillo s/n, 28692 Villanueva de la Ca\~nada, Madrid, Spain\label{cab}
\and Leiden Observatory, Leiden University, P.O. Box 9513, 2300 RA Leiden, the Netherlands\label{leiden}
\and Department of Astronomy, Oskar Klein Centre, Stockholm University, 106 91 Stockholm, Sweden\label{oskar}
\and School of Cosmic Physics, Dublin Institute for Advanced Studies, 31 Fitzwilliam Place, Dublin, D02 XF86, Ireland\label{dublin}
\and Jet Propulsion Laboratory, California Institute of Technology, Pasadena, California, United States\label{jpl}
}

 \abstract
   {The MIRI instrument onboard JWST is now offering high-contrast imaging capacity at mid-IR wavelengths, thereby opening a completely new field of investigation to characterize young exoplanetary systems.}
   {The multiplanet system HR\,8799 is the first target observed with MIRI's coronagraph as part of the MIRI-EC Guaranteed Time Observations exoplanets programme, in Nov. 2022. We obtained deep observations in three coronagraphic filters from $\sim 10$ to 15\,$\muup$m (F1065C, F1140C, F1550C), and one standard imaging filter at $\sim 20\,\muup$m (F2100W), with the goal to extract the photometry of the four planets, as well as to detect and investigate the distribution of circumstellar dust.}
   {Using dedicated observations of a reference star, we tested several algorithms to subtract the stellar diffraction pattern while preserving the fluxes of planets, which can be significantly affected by over-subtraction. Measuring correctly the planet's flux values requires accounting for the attenuation by the coronagraphs as a function of their position, and to estimate the normalisation with respect to the central star. We tested several procedures to derive averaged photometric values and error bars.}
   {These observations have enabled us to obtain two main results.   
   First of all, the four planets in the system are well recovered, and their mid-IR fluxes, combined with near-IR flux values from the literature, are compared to two exoplanet atmosphere models, \texttt{ATMO} and \texttt{Exo-REM}. As a main outcome, the MIRI photometric data points imply larger radii (0.86 or 1.07\,$R_\mathrm{J}$ for planet b) and cooler temperatures (950 or 1100\,K for planet b), especially for planet b, in better agreement with evolutionary models.   
   Second of all, these JWST/MIRI coronagraphic data also deliver the first spatially resolved detection of the inner warm debris disk, the radius of which is constrained to about 15\,au, with flux densities comparable, but lower than former unresolved spectroscopic measurements with Spitzer. }
   {The coronagraphs of MIRI cast a new vision of known exoplanetary systems which differs significantly from more shorter wavelength high-contrast images delivered by extreme adaptive optics from the ground. Inner dust belts and background galaxies become predominant at some mid-IR wavelengths, potentially causing confusion for the detection of exoplanets. Future observing strategy and data reduction should take such features into account.    }
\authorrunning{A. Boccaletti et al.}
\titlerunning{Imaging the HR 8799 system with JWST’s MIRI coronagraph}

\date{submitted to A\&A on Sept. 8th, 2023 }

\maketitle


\begin{table*}[th!]
\begin{center}
\caption{Main parameters of the observations during JWST/MIRI program 1194: date/time, order of execution in the sequence, filter, name of the object, type of object (target or reference, or background image), identifier of the sequence, number of groups, number of integrations, number of dither positions (9 is for the SGD, 4 or 1 is a classical dither, 2 is for the background), total exposure time per dither.} 
\label{tab:log}

\begin{tabular}{llllllllll}
\hline \hline
date / time in UT  & seq. & filter & object   &  type     & obs id  & N$_{group}$ & N$_{int}$ & N$_{dither}$ & T$_\mathrm{exp}$                          \\
(month/day/year)          &       &        &          &           &         &             &           &        &  per dither (s) \\  
\hline
11/07/2022 21:19 & 1 & F1065C & HR\,8799     & TARG        & obs 7  & 500 & 9   & 1 & 1080.48 \\
11/07/2022 22:43 & 3 & F1065C & $-$          & BGD           & obs 13 & 500 & 9   & 2 & 1080.48 \\
11/07/2022 23:49 & 4 & F1065C & HD\,218261   & REF         & obs 14 & 500 & 2   & 9 & 239.92 \\
11/08/2022 01:42 & 6 & F1065C & $-$          & BGD           & obs 20 & 500 & 2   & 2 & 239.92 \\
\hline
11/08/2022 08:24 & 7 & F1140C & HR\,8799     & TARG        & obs 8  & 500 & 9   & 1 & 1080.48 \\
11/08/2022 13:25 & 10 & F1140C & $-$          & BGD           & obs 12 & 500 & 9   & 2 & 1080.48 \\
11/08/2022 14:27 & 11 & F1140C & HD\,218261   & REF         & obs 15 & 500 & 2   & 9 & 239.92 \\
11/08/2022 16:42 & 14 & F1140C & $-$          & BGD           & obs 19 & 500 & 2   & 2 & 239.92 \\
\hline
11/08/2022 08:57 & 8 & F1550C & HR\,8799     & TARG        & obs 9  & 500 & 36  & 1 & 4322.63 \\
11/08/2022 10:43 & 9 & F1550C & $-$          & BGD           & obs 11 & 500 & 36  & 2 & 4322.63 \\
11/08/2022 15:28 & 12 & F1550C & HD\,218261   & REF         & obs 16 & 500 & 2   & 9 & 239.92 \\
11/08/2022 16:27 & 13 & F1550C & $-$          & BGD           & obs 18 & 500 & 2   & 2 & 239.92 \\
\hline
11/07/2022 22:06 & 2 & F2100W & HR\,8799     & TARG        & obs 10 & 20  & 50   & 4 & 314.20 \\
11/08/2022 00:50 & 5 & F2100W & HD\,218261   & REF         & obs 17 & 20  & 05   & 4 & 314.20 \\
\hline
\end{tabular}
\end{center}
\end{table*}

\section{Introduction}
\label{sec:introduction}

Massive giant planets in large orbits ($>$5\,au) 
are found to be relatively rare, as inferred from direct imaging surveys \citep{vigan_sphere_2020, nielsen_gemini_2019}, and even the statistics from radial velocity surveys are not complete in that case \citep{lagrange_radial_2023}.  
Therefore, every single detection is significant to add to our understanding of how such massive planets can form in the outer parts of planetary systems, and how they can impact the fate of other planets. Historically, young giant planets have been discovered and characterized in the near-IR with general facilities, or dedicated high-contrast instruments on the ground. But, little is known about their properties at longer wavelengths. Observations from the ground at mid-IR have proven to be difficult because of sensitivity issues and essentially yield no clear detection \citep{wagner_imaging_2021,skaf_-sky_2022}. The James Webb Space Telescope (JWST) mission \citep{gardner_james_2023} is definitely a game changer in this field by providing high-angular resolution together with high-contrast capacities at wavelengths longer than $\sim5\,\muup$m that were not even accessible before. 
In this respect, mid-IR is crucial to put independent constraints on planets luminosity, temperature, and radius, as well as to provide access to molecules like ammonia. \cite{carter_jwst_2023} presented JWST observations of a giant planet, previously discovered with SPHERE at the VLT \citep{chauvin_discovery_2017}, and observed for the very first time at 10 and 15\,$\muup$m with the JWST Mid-Infrared Instrument \citep[MIRI,][]{wright_mid-infrared_2015} and its coronagraphic system \citep{boccaletti_mid-infrared_2015}.

Since its discovery in 2008 at Keck \citep{marois_direct_2008, marois_images_2010}, the HR\,8799 multiplanet system has been the focus of numerous studies to explore atmospheric properties, and dynamics, as well as the dust content of its debris disk. Consisting of four giant planets with masses ranging roughly between 5 and $10\,M_{J}$ according to hot start evolutionary models, the system is often seen as an upscale version of a young Solar System \citep{faramaz_detailed_2021}. 
The HR\,8799 system is very unique among all known exoplanetary systems as it provides a testbed for understanding the formation of planets around early-type stars.

With the objective to characterize the atmospheres of these four planets, spectral analysis has been performed for a broad range of near-IR wavelengths from the {\it z} to M bands, in photometry \citep{currie_combined_2011, skemer_directly_2014}, at low spectral resolution \citep{bonnefoy_first_2016, ingraham_gemini_2014, lacour_first_2019} and even with medium to high resolution \citep{barman_clouds_2011, konopacky_detection_2013, ruffio_deep_2021, wang_atmospheric_2022}. 
%
Their colors indicate that these four giant planets differ from field brown dwarfs with notably redder colors, near the L/T transition, which require patchy optically thick clouds and non-equilibrium chemistry to match the observations. Their temperature range from 900\,K to 1300\,K, with planet b being significantly cooler and fainter, while surface gravity ranges between $\mathrm{log}(g)=3.5$ and $4.5$ \citep{bonnefoy_first_2016}. 
These near-IR data suggest a tendency towards super-solar metallicity. While atmospheric models reproduce rather well the near-IR photometry and spectrum for planets d and e, they are much less effective in fitting planets b and c \citep{bonnefoy_first_2016}. One issue has been the determination of planet radii which are found to be too small, in particular for planet b with values as small as $0.5-0.7\,R_{J}$. For an assumed age of  about 30\,Myrs evolutionary models predict radius which can be as large $1.3\,R_J$.

At least two molecules are clearly identified, namely H$_2$O and CO, even at low spectral resolution, while there is a debate about the detection of methane, which is not reproducible from one data reduction to the other \citep{barman_simultaneous_2015, petit_dit_de_la_roche_molecule_2018}. This is one main difference with field brown dwarfs, the spectra of which show the signature of methane at temperatures cooler than $\sim$\,1300\,K. In addition, the measurement of the C/O ratio is claimed as one of the most promising ways to constrain the formation history \citep{oberg_effects_2011}, and to disentangle between accretion-like and stellar-like mechanisms, the former scenario predicting super-solar values. As of today, there is no clear consensus on the C/O ratio since various studies yield solar, or marginally super-solar values \citep{molliere_retrieving_2020, konopacky_detection_2013, ruffio_deep_2021}.
Some of the atmospheric properties are well explained by self-consistent atmospheric models like \texttt{Exo-REM} \citep{charnay_self-consistent_2018}. In particular, the colors with respect to the L/T transition can be explained by low  gravity planets with clouds, although there is an alternative interpretation with cloudless models but involving fingering convection \citep{tremblin_cloudless_2016}. \cite{charnay_self-consistent_2018} also explained how low gravity and clouds contribute to reducing the abundance of methane.

The HR\,8799 multiplanet system also contains planetesimals belts which are replenishing the system with small dust particles in a collisional cascade. The dust architecture has been essentially inferred  from {Spitzer} spectroscopy \citep{chen_spitzer_2006}, and  IR excess analysis, from which \citet{su_debris_2009} derived the presence of  three components: an inner warm ($\sim$\,150\,K) belt located at $\sim$\,6 to 15\,au, an outer cold ($\sim$\,45\,K) belt at $\sim$\,90 to 300\,au, and a halo further out. So far, only the outer cold belt has been spatially resolved at sub-millimeter and millimeter wavelengths \citep{hughes_resolved_2011, booth_resolving_2016, faramaz_detailed_2021}. In this picture, the outer edge of the inner belt, and the inner edge of the outer belt, would be sculpted by planet e and b, respectively. However, the size of the disk is not fully in agreement with the analysis of the IR excess which raises some debate about the presence of a fifth planet \citep{faramaz_detailed_2021}. 

As of today, the HR\,8799 multiplanet system has been observed extensively in the near-IR and far-IR regimes, which are complementary and relevant to investigate different physical processes, but was lacking any deep exploration in the mid-IR \citep{petit_dit_de_la_roche_new_2020}, which is interesting in the context of planet formation. 
As to atmosphere characterisation, mid-IR observations at wavelengths longer than 5\,$\muup$m are valuable to provide more direct measurements of the effective temperature and radius of giant planets with less degeneracies caused by clouds, as well as to discriminate between various atmosphere models. Mid-IR observations encompass the signature of ammonia which has some advantage over methane to provide temperature estimates at least for $T<1000\,K$ \citep{danielski_atmospheric_2018}. Moreover, the broad signature of silicate particles can constrain the composition of clouds if present \citep{miles_jwst_2022}. Additionally, this spectral range matches the peak of the emission of planets cooler than those with a peak flux in the near-IR, providing means to push detections to lower masses. Finally, the mid-IR has the ability to probe the intermediate-size grain population (a few to tens of $\muup$m) which can be distributed differently than sub-micron sized grains, seen in scattered light, and the larger millimeter sized grains.

In this paper we present the first spatially resolved detection of the four planets in the HR\,8799 system, and of the inner warm disk, in the mid-infrared, with JWST/MIRI. Sect. \ref{sec:datared_obs} describes the observations and the data reduction. We detail the photometric measurements in Sect. \ref{sec:photom} which are used to perform the atmospheric modeling (Sect. \ref{sec:atmosphere}). The detection of the inner disk is addressed in Sect. \ref{sec:disk}. Finally, the nature of a background object is discussed in Sect. \ref{sec:galaxy}, and we conclude in Sect. \ref{sec:conclusion}.

\section{Observations and data reduction}
\label{sec:datared_obs}

\subsection{Observations}
\label{sec:observations}


HR\,8799 was observed in two runs on Nov. 7th, 2022 (filters F1065C and F2100W), and Nov. 8th, 2022 (filters F1140C and F1550C), under GTO (Guaranteed Time Observations) program 1194, using both MIRI's Four Quadrant Phase Mask (4QPM) coronagraphs, and standard imaging. The log of observations is provided in Table \ref{tab:log}. 
For each coronagraphic filter, we observed back-to-back the target and its associated background (in two dithers), and then the reference star together with its own background. Background images are necessary to remove the "glowstick" effect, as explained in \citet{boccaletti_jwstmiri_2022}, and are obtained near the target (typically a few tens of arcseconds away). On the contrary, no background observation is necessary for non-coronagraphic imaging data since the star is dithered in four positions across the field of view.

The reference star, HD/,218261, was chosen 1) to be angularly close to the target ($\sim1.24\degb$) to minimize the thermal drift induced by wavefront errors, and 2) with comparable magnitudes to HR\,8799 in both the K band, and the MIRI's coronagraphic filters, so that it can be shared with the NIRCAM GTO program, with the goal to optimize overheads in telescope pointing. Taking into account the stellar residuals and the background noise from our diffraction model \citep{boccaletti_mid-infrared_2015}, the exposure times were determined to achieve signal-to-noise ratios (S/N) larger than $\sim10$ on the planets. The total exposure times on target are  1080\,s, 1080\,s, 4322\,s, and  1257\,s for the F1065C, F1140C, F1550C, and F2100W filter respectively.  

\begin{figure*}[ht]
    \centering
    \includegraphics[width=18cm]{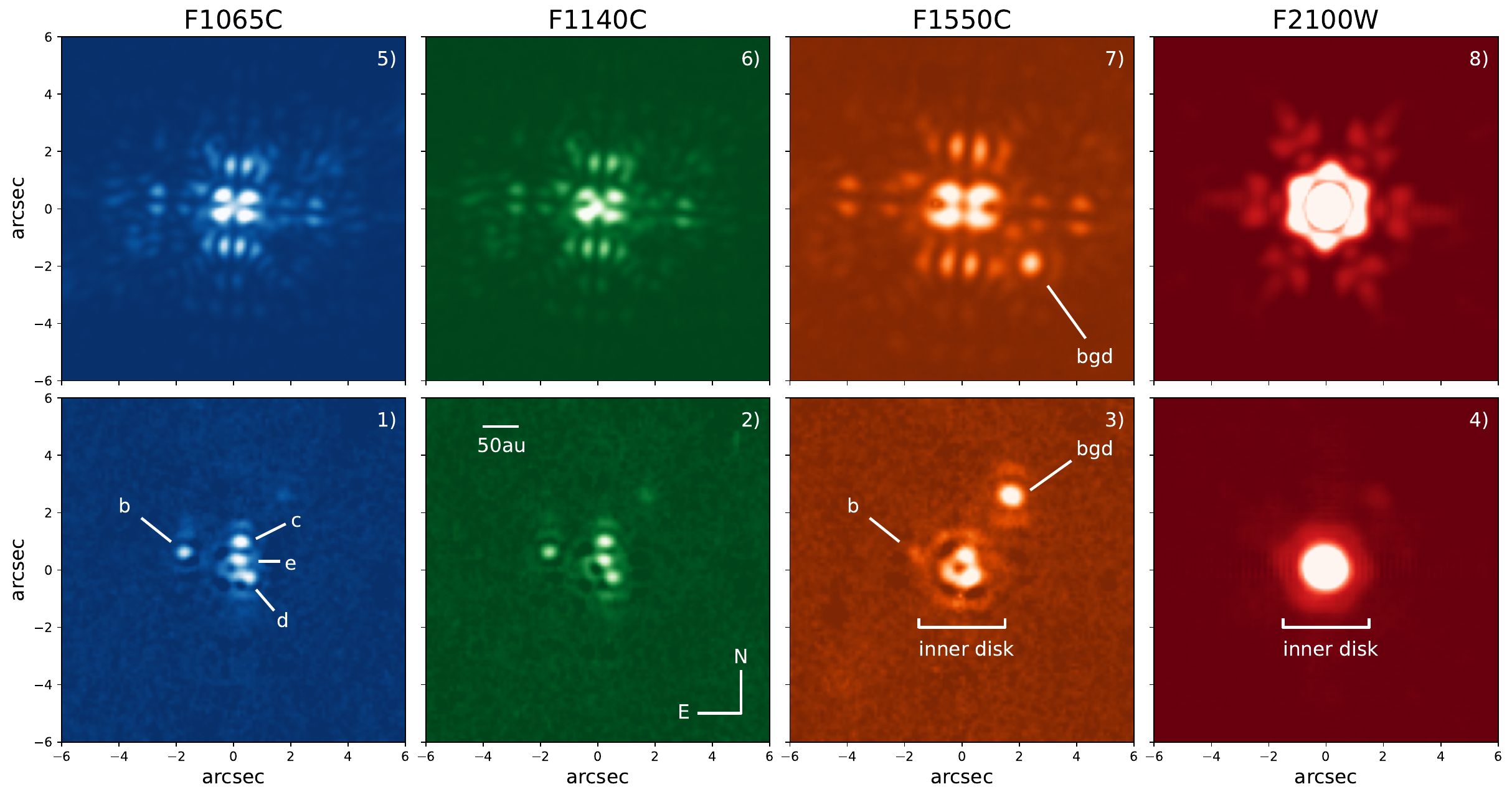}
    \caption{Raw coronagraphic (top, detector orientation), and reference star subtracted (bottom, North is up, East is left) images in the four filters (left to right: F1065C, F1140C, F1550C, F2100W). The four planets are labelled b, c, d, and e in panel 1), as well as the background object in panels 3) and 7). The signature of the inner disk is circled with a dotted line in panels 3) and 4). The 50\,au scale stands for the projected physical distance. The Field of View is $12\times 12"$. Intensity scale is adapted in each panel for visualisation purpose.}
    \label{fig:raw_and_corono_images}
\end{figure*}

\subsection{Data reduction}
\label{sec:datared}

We retrieved the processed data from the Mikulski Archive for Space Telescopes (MAST\footnote{\url{https://mast.stsci.edu/portal/Mashup/Clients/Mast/Portal.html}}), and also reprocessed the raw data with our own routines for comparison. 
We outline here the main steps of the process. 
Stage 1 of the {JWST} pipeline\footnote{\href{https://jwst-pipeline.readthedocs.io/en/latest/}{jwst-pipeline.readthedocs.io}} is taking as input the uncalibrated files (\texttt{uncal}) and applies the detector-level corrections to raw JWST ramps to produce the 2D slope product. Stage 1 corrects for the dark current, and bad pixels such as those that would be saturated or impacted by cosmic rays that would produce jumps in the ramp. \cite{carter_jwst_2023} noted that the default jump threshold value  is too low and leads to erroneously flagged bad pixels. Therefore, we ran the steps with different values, but we find that it does not have any impact on the calibration. 
In comparison to ERS data \citep{hinkley_jwst_2022}, the ramps in this program have many more groups per integration. Therefore, the jumps have less impact on the calibrated slopes. The first and last group of each integration are removed, and the ramps are fitted to generate the \texttt{rate} files. Stage 2 of the {JWST} pipeline is meant to  photometrically calibrate the data. However, because the 4QPM produces a shadow pattern along its transition due to diffraction effects, we skip the \texttt{flat$\_$field} steps to avoid increasing the noise as well as the "glow-stick" effect close to the mask center (precisely where we expect to detect the planets). The photometric calibration in the pipeline is also skipped, as long as we apply our own photometric calibration, as detailed in the Sect. \ref{sec:extracphotom}. 
In practice, stage 2 is only taking care of the background subtraction. 

In the end we made use of the \texttt{rate} and \texttt{cal} files which collapse all integrations (N$_{int}$) in a single frame, yielding a number of frames that is the number of dither positions (N$_{dither}$). 
All data reductions agree visually in terms of planet detection as well as photometrically. 

A mean combination of the background dithers is subtracted from each target observation, and we further reject the remaining bad pixels with sigma clipping. 
Each coronagraphic image is registered at the 4QPM centers which was determined during commissioning (using cross-correlation with a large database of simulated data). 

Raw coronagraphic images are essentially dominated by the diffraction. This specific signature shown in Fig. \ref{fig:raw_and_corono_images}\,(top) is a result of the particular JWST hexagonal pupil, that even an optimized Lyot stop cannot completely suppress \citep{boccaletti_imaging_2005}. The classical workaround is to observe a reference star in similar conditions referred to as Reference Differential Imaging (RDI). But, it has been shown that the contrast achieved with MIRI’s 4QPM is mostly limited by the relative pointing accuracy between the target star and the reference star \citep{cavarroc_target_2008}, which is estimated to about $5-10$\,mas \citep{rigby_science_2022}. To mitigate this issue, \citet{soummer_small-grid_2014} and \citet{lajoie_small-grid_2016} proposed to implement the Small Grid Dither (SGD) strategy. While we obtain a single pointing of the target (hence with a precision of about $5-10$\,mas), several (5 or 9) pointings of the reference star are performed to provide diversity, and to match, ideally, the diffraction pattern of the reference star with that of the star. Elaborate algorithms like \texttt{PCA} \citep{soummer_detection_2012} and \texttt{LOCI} \citep{lafreniere_new_2007} can provide an optimal subtraction of this diffraction pattern. 

For the coronagraphic observations of HR\,8799, we used the 9-point SGD, defined as a regular square grid with 10\,mas steps. We compared several types of algorithms to suppress the star's diffraction pattern, but since the diversity is low, the best results are obtained with a linear combination of the nine reference frames, of which the coefficients are calculated with a downhill simplex method (referred to as \texttt{amoeba}) applied to a restricted field of view of 0.4$''$ to 8$''$ in radius (Fig. \ref{fig:raw_and_corono_images},\,bottom). The frames combination which minimizes the residuals in this field is then subtracted out from the target image.

The so-called over-subtraction \citep{pueyo_detection_2016} can be particularly strong in the case of MIRI observations given that the angular separations of the planets are not significantly larger with respect to the angular resolution (about 0.3$''$ at 10\,$\muup$m). 
To overcome this issue, 
we further masked the planets in the RDI process. In practice, we used a simple patch of 1\,$\lambda/D$ in radius to remove the contribution of the planet's signal in the minimization of the residuals. We also tested a method which subtracts the planet's diffraction pattern (taking into account the coronagraph's transmission, see Sect. \ref{sec:corotrans}), but this requires a first iteration with masks to evaluate the planets' fluxes. Both methods were found to be qualitatively similar although the latter provides larger error bars. 
Figure \ref{fig:raw_and_corono_images} displays the results with the masking technique, while other methods including variations of \texttt{amoeba}, together with a median combination of the reference frames, and \texttt{PCA} with various sets of reference stars, are presented in Figure \ref{fig:allalgo} in the appendix. 

For the observations in the F2100W filter in standard imaging mode, we used a standard 4-point dithering to minimize the impact of the background and cosmetics of the detector. The reference star is observed the same way. Since the inner part of the image (essentially the PSF peak) is dominated by the inner disk flux (see sec. \ref{sec:disk}), we performed a direct weighted subtraction to minimize the diffraction in the $2-5''$ annular region (to exclude the region where the inner disk dominates). 

\subsection{Overall description of the images}

The four known planets, HR\,8799 bcde, are well detected at the two shortest wavelengths (F1065C in Fig. \ref{fig:raw_and_corono_images}-1, and F1140C in \ref{fig:raw_and_corono_images}-2), while only planet b is detected at F1550C (Fig. \ref{fig:raw_and_corono_images}-3). 
However, interestingly, planet c is also barely detected at F1550C with the \texttt{PCA} reduction technique using the commissioning stars as references (Fig. \ref{fig:allalgo}).   
The longest wavelength filter (F2100W, Fig. \ref{fig:raw_and_corono_images}-4) yields no planet detection. Another point source to the northwest of the star, a suspected background galaxy \citep{faramaz_detailed_2021} labeled "bgd" in Fig. \ref{fig:raw_and_corono_images}-3 and Fig. \ref{fig:raw_and_corono_images}-7, appears to be very bright at F1550C, and is also visible, but much fainter, in the other three filters (see Sect. \ref{sec:galaxy}). Finally, at the longest wavelengths (F1550C in Fig. \ref{fig:raw_and_corono_images}-3, and F2100W in \ref{fig:raw_and_corono_images}-4), residual emission at the center of the images corresponds, presumably, to the inner warm debris disk. The same feature is also identified at shorter wavelengths with much lower S/N ratios. The doughnut-like shape of the inner disk is due to the coronagraphic effect, and the dashed circle in  Fig. \ref{fig:raw_and_corono_images}-3 is not representing the actual size of the disk but the extent of its image (see Sect. \ref{sec:disk} for details). 
Overall, the MIRI images of the HR\,8799 system yield a very different vision than in the near IR, with the clear detection of the four planets, together with a localized but extended central emission.

\section{Photometry of the planets}
\label{sec:photom}

Extracting the photometry in the MIRI coronagraphic images requires normalization of the planet's flux to the stellar Point Spread Function (PSF) to derive the contrast values, as well as to take into account the planet's attenuation due to the coronagraph and to estimate photometric error bars. 

\subsection{PSF normalization}
\label{sec:psfnorm}

To avoid saturation, the star cannot be observed out of the coronagraphic mask. There is no direct way to measure the star-to-planet contrast in the very same filter with this program, and this is certainly a limitation for accurate photometry. Instead, we considered two solutions for normalizing the stellar flux, both relying on commissioning (COM) data for which we manage to observe a PSF (out-of-mask image) and a coronagraphic image (on-mask image). 
The first solution makes use of the target acquisition data (TA) in addition to the COM data. TA comes with any coronagraphic sequence to estimate the centroid of the target in a coronagraph subarray in order to move the star right at the location of the coronagraph. 
The second solution involves the COM data alone which was designed to accurately measure contrasts for all filters, carefully choosing a star that is not too bright to obtain unsaturated PSFs \citep{boccaletti_jwstmiri_2022}.

For TA normalization (Eq. \ref{Eq:norm_ta}), the method consists in measuring the flux ratio of the TA images between the target, HR\,8799 ($I_{TA\_targ}$), and the commissioning star ($I_{TA\_com}$), and to apply this factor to the commissioning PSF ($I_{PSF\_com}$) to generate a pseudo HR\,8799 PSF ($I_{PSF\_targ}^{TA}$), as follows : 

\begin{equation}
     I_{PSF\_targ}^{TA} = I_{PSF\_com}\times\ \dfrac{\sum\limits_{k}A_k.I_{TA\_targ}}{\sum\limits_{k}A_k.I_{TA\_com}}
     \label{Eq:norm_ta}
\end{equation}

$A_k$ is a circular aperture containing $k$ pixels which is set here to a 1$''$ radius. The two TA images (at two dithered positions) follow the same data reduction procedure as other data. We keep the one that is further out from the 4QPM center to avoid attenuation effects after subtracting them to get rid of the background. 
In the case of HR\,8799 observations, TA data were obtained with the neutral density filter (FND). The FND is intentionally broad ($8-18\,\muup$m) to mitigate the chromatic effects due to the 4QPM, avoiding bias to the centroid estimation \citep{cavarroc_first_2008}. But, normalization issues can arise if the science target has a different spectral slope in the FND spectral range than the commissioning star. This is particularly problematic when IR emission, like an unresolved (or slightly resolved) disk, adds to the total flux of the central source. At the moment, only the FND has been validated against the TA procedure, but for future programs MIRI will allow TA to be carried out with F560W, F1000W, and F1500W filters too. 

The second solution is based on the calibration of the coronagraph attenuation on the central star using commissionning data. Here, the normalization factor (Eq. \ref{Eq:norm_coro}) corresponds to the flux ratio of the coronagraphic images ($I_{CORO\_targ}$, and $I_{CORO\_com}$). Because, these images are much more extended than in the former case, $A_k$ is representing an aperture of 5$''$ in radius to encompass the full coronagraphic diffraction pattern. As an advantage with respect to the former solution, this method uses the same coronagraphic filter. Still, the target and commissionning star are different, and again, the presence of an unresolved inner disk can bias the normalization. This method is also prone to biasing because of background objects since the aperture, $A_k$, is much larger to encompass the full coronagraph diffraction pattern. The pseudo-PSF reads:

\begin{equation}
     I_{PSF\_targ}^{CORO} = I_{PSF\_com}\times\ \dfrac{\sum\limits_{k}A_k.I_{CORO\_targ}}{\sum\limits_{k}A_k.I_{CORO\_com}}
     \label{Eq:norm_coro}
\end{equation}

Since we have two dithers for $I_{PSF\_com}$ (rejecting two other dithers for image quality reasons), and nine dithers of $I_{CORO\_com}$ (because observations used 9-point SDG) we can evaluate two, respectively eighteen, values of the star's count, for $I_{PSF\_targ}^{TA}$, respectively $I_{PSF\_targ}^{CORO}$, contributing to the uncertainties in the planets' photometry. Therefore, the estimate of the dispersion for each method ($TA$ and $CORO$) is not based on the same amount of independent realizations. 
But overall, we estimate that the normalization factors obtained with the 
two methods are consistent within 5\%, 16\% and 37\% for filters F1065C, F1440C, F1550C, respectively. The larger discrepancy at F1550C can be explained by the background contamination, which is stronger at longer wavelengths, and can bias the $CORO$ method. The background galaxy, easily identifiable in the raw coronagraphic frame of HR\,8799 (Fig. \ref{fig:raw_and_corono_images}-7), is one potential source of bias.   

\begin{table}[t]
    \centering
        \caption{
        Transmission of a point source for the three MIRI 4QPM coronagraphs at each planet's location calculated either in an aperture region of 1.5\,$\lambda/D$ in radius, or integrated in the whole image.}

    \begin{tabular}{llll} \hline \hline
        planet  & F1065C    & F1140C    & F1550C    \\ \hline 
        b       &   0.79 / 0.82 &   0.75 / 0.79    &   0.76 / 0.82    \\
        c       &   0.77 / 0.80 &   0.74 / 0.78    &   0.61 / 0.65    \\
        d       &   0.61 / 0.67 &   0.59 / 0.65    &   0.62 / 0.66    \\
        e       &   0.56 / 0.58 &   0.52 / 0.56    &   0.40 / 0.45    \\ \hline 
    \end{tabular}
    \label{tab:transmission}
\end{table}

\subsection{Coronagraphic transmission}
\label{sec:corotrans}

Each planet in the image  has a transmission which depends on its position with respect to the 4QPM center and phase transitions, and which also scales with wavelength. We estimate this transmission using the diffraction model from \citet{boccaletti_mid-infrared_2015} calibrated against commissioning data \citep{boccaletti_jwstmiri_2022}, with updates of the telescope wavefront error measurements. This diffraction model is based on a similar concept as \texttt{webbpsf} \citep{oschmann_updated_2014}, so that the planet's transmission can be also evaluated with the latter. 

We used \texttt{whereistheplanet} \citep{wang_whereistheplanet_2021} to estimate the planets' locations at the epochs of observations with respect to the 4QPM mask orientation (which is inclined by 4.835$\degb$ with respect to the detector), accounting for the field orientation (the V3PA parameter in the JWST coordinate system). The estimated values of the coronagraphs' transmission are provided in Table \ref{tab:transmission} and used to correct for the planets' photometry.
To give an idea of the impact of the coronagraph transmission, we display in Fig. \ref{fig:planetsalone}  the image of four point sources of equal intensities located at the positions of the  HR8799 planets.

\subsection{Planets' contrasts and fluxes}
\label{sec:extracphotom}

The planets' photometry is assessed in two different ways. First,  we integrate the planets' count rates (in DN/s) in RDI images, in an aperture of 1.5\,$\lambda/D$ in radius, the size of which is found to be optimal to integrate enough planets' signal while keeping the contamination of the other planets at a low level. Second, we use negative fake planets. As explained in section \ref{sec:corotrans}, we modeled the planets' diffraction pattern for each filter and each planet positions with respect to the 4QPM transitions. The minimization of the residuals in the  1.5\,$\lambda/D$ area accounts for three parameters: the flux and the positions of the planet's model (which is allowed to vary by a few pixels compared to the theoretical positions). The final contrast is the count rates measured in the PSF (Sec. \ref{sec:psfnorm}) relative to the count rates measured in the planet, in the same patch, and corrected by the local attenuation due to the coronagraph. With respect to the compensation of the inherent over-subtraction of RDI mentioned in Section \ref{sec:datared} we opt for the method using masks which provides lower error bars. 
Irrespective of the PSF normalization, the two methods to extract planet's photometry agree within 3\% to 46\% depending on filters and planets, in some cases resulting in two families of contrast values (for instance for planet b in F1550C, see Table \ref{tab:contrasts}). Table \ref{tab:contrasts_mean} provides the average contrast values and standard deviations together with the dispersion between the two extraction methods, while the individual measurements for each PSF normalization and flux extraction methods are given in the Appendix (Tab. \ref{tab:contrasts}).

\begin{table}[t]
    \centering
        \caption{Star-to-planet contrasts measured for planets b, c, d, and e, in the three coronagraphic filters. The values in brackets correspond to the dispersion due to the flux extraction method (aperture and negative fake planets). See Sect. \ref{sec:extracphotom} for details.}
    \begin{tabular}{llll}\hline \hline

       planet       &   F1065C          &   F1140C          &   F1550C          \\ \hline        b            &   $2609\pm104$    &   $2375\pm154$    &   $2131\pm463$    \\
                    &   [3\%]           &   [7\%]           &   [39\%]          \\
       c            &   $1301\pm172$    &   $1316\pm213$    &   $>489$          \\
                    &   [25\%]          &   [30\%]          &                   \\
       d            &   $1295\pm283$    &   $1292\pm221$    &   $>385$          \\
                    &   [38\%]          &   [31\%]          &                   \\ 
       e            &   $885\pm208$     &   $878\pm186$     &   $>242$          \\ 
                    &   [46\%]          &   [40\%]          &                   \\  \hline 
    \end{tabular}
    \label{tab:contrasts_mean}
\end{table}

\begin{figure*}
    \centering
    \includegraphics[width=18cm]{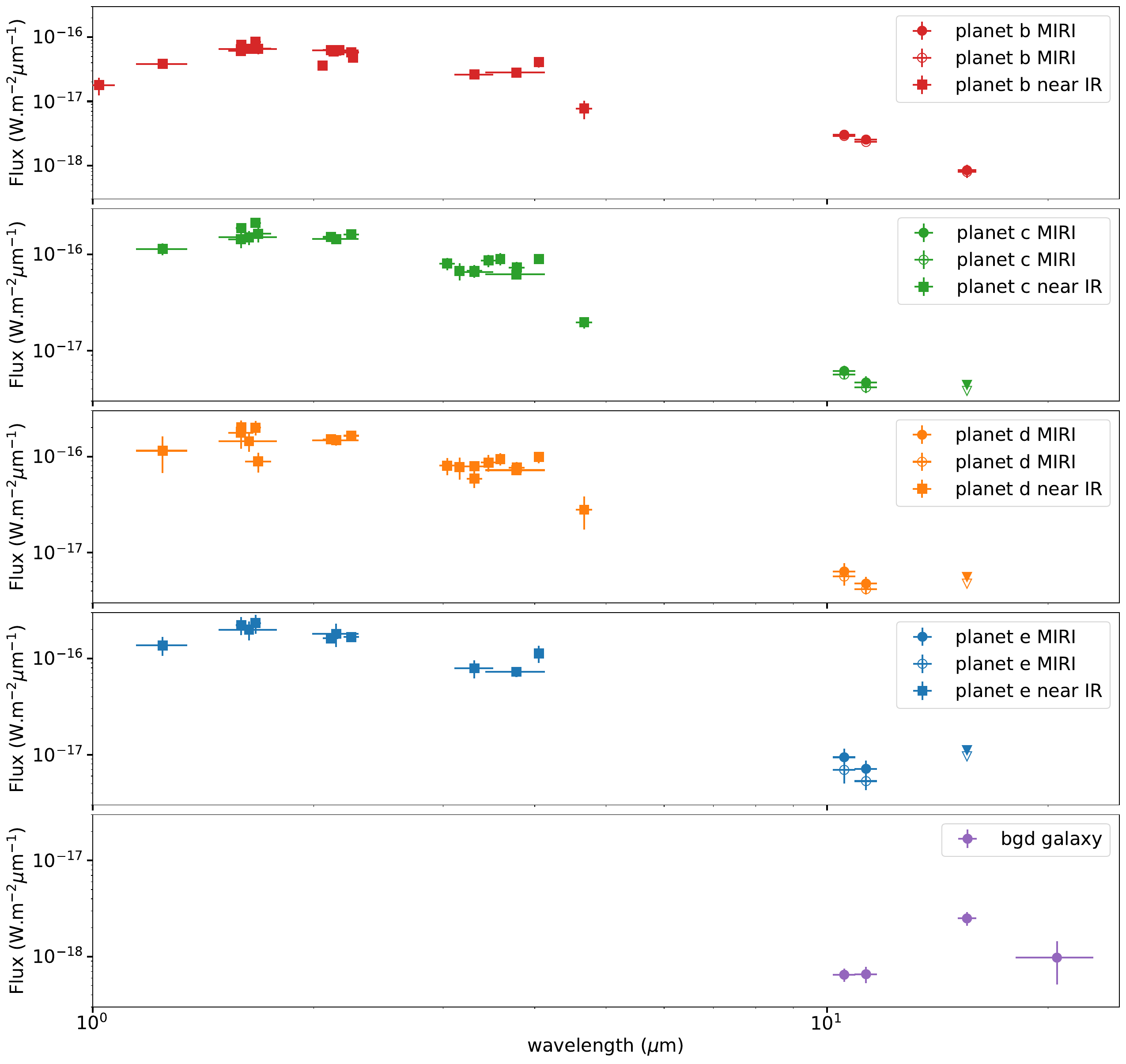}
    \caption{Flux density in W.m$^{-2}.\muup$m$^{-1}$ of the four planets and the background object in the MIRI filters, with (filled circles) and without (unfilled circles) correcting for the over-subtraction, and in the near IR (squares). Triangles denote flux upper limit. }
    \label{fig:allplanets}
\end{figure*}

Converting contrasts to fluxes requires a stellar flux model. 
We retrieved synthetic photometry from VOSA \citep[Virtual Observatory SED analyzer,][]{bayo_vosa_2008} considering the BT NextGen stellar model \citep{allard_model_2011, asplund_chemical_2009} with an effective temperature T$_{eff}$=7600\,K, a surface gravity $\mathrm{log}(g) = 4.5$\, cm.s$^{-2}$ , and solar metallicity. We adopt a stellar radius of 1.34\,R$_{\odot}$, and a distance $d=40.88$\,pc \citep{gaia_collaboration_gaia_2021}. 
Further, we obtained actual photometry of HR\,8799 from VizieR in the 2MASS and WISE filters: J, H, Ks, W1, W2 and W3 (Table \ref{tab:starphotom}), excluding the shorter wavelengths which may not be representative to interpolate the mid IR fluxes, as well as wavelengths longer than 20\,$\muup$m to avoid being biased by the emission from the debris disk. We perform a $\chi^2$ minimization to determine the global intensity offset between the model and the real star's photometric data, and find a correction of a factor of 1.22. Figure \ref{fig:allplanets} shows the flux density of the four planets as measured in the MIRI filters, with the over-subtraction compensated (and without to see the corresponding impact), together with the near-IR photometry from the literature, which is compiled in \citet{bonnefoy_first_2016}. The flux densities are reported in Table \ref{tab:planet_photom} for each planet and each filter. 

\begin{table}[t]
    \centering
        \caption{Photometry of the star retrieved from VizieR.}
    \begin{tabular}{lllll}\hline \hline
    wavelength 	& flux          &	flux  &	flux error   &	filter \\ 
     ($\muup$m) &  (W.m$^{-2}$)   & (Jy)    &    (Jy)      &     \\ \hline 
    1.24		& 2.69e-11	& 11.1		& 0.3		& 2MASS:J\\
    1.65		& 1.47e-11	& 8.11		& 0.13		& 2MASS:H\\
    2.16		& 7.50e-12	& 5.41		& 0.09		& 2MASS:Ks\\
    3.35		& 2.29e-12	& 2.56		& 0.52		& WISE:W1\\
    4.60		& 1.07e-12	& 1.64		& 0.12		& WISE:W2\\
    11.6		& 6.17e-14 	& 0.238		& 0.004		& WISE:W3    \\ \hline
    \end{tabular}
    \label{tab:starphotom}
\end{table}

\section{Atmospheric modelling }
\label{sec:atmosphere}
We explored the properties of the planets' atmospheres with two complementary models : \texttt{Exo-REM}  and \texttt{ATMO}, both developed for long period young giant planets, neglecting the stellar irradiation. They both provide grids of synthetic spectra with various atmospheric parameters. 

\begin{table}[ht]
    \centering
        \caption{Flux density ($\times 10^{-18}$) in W.m$^{-2}$.$\muup$m$^{-1}$ of the four planets in the MIRI filters (min/max values).}

    \begin{tabular}{llll}\hline \hline
       planet       &   F1065C          &   F1140C         &   F1550C          \\ \hline 
       b            &   2.90 / 3.14     &   2.37 / 2.70    &   0.66 / 1.03     \\
       c            &   5.34 / 6.97     &   3.92 / 5.43    &   $<$ 2.90        \\
       d            &   4.98 / 7.78     &   3.96 / 5.60    &   $<$ 3.69        \\
       e            &   7.20 / 11.63     &  5.63 / 8.65    &   $<$ 5.09       \\ \hline
    \end{tabular}
    \label{tab:planet_photom}
\end{table}

\texttt{Exo-REM} \citep{baudino_interpreting_2015, charnay_self-consistent_2018} is a self-consistent 1D radiative-convective equilibrium model which assumes the net flux is conserved. It incorporates the opacities of collision-induced absorptions (H$_2$–H$_2$, H$_2$–He, H$_2$O-H$_2$O and H$_2$O–N$_2$), and  rovibrational bands from various molecules (H$_2$O, CH$_4$, CO, CO$_2$, NH$_3$, PH$_3$, TiO, VO, H$_2$S, HCN, and FeH), as well as resonant lines from sodium and potassium. \texttt{Exo-REM} uses micro-physics to model clouds of silicate, iron, sulfide, alkali salt, and water clouds, and it handles disequilibrium chemistry as well.

\texttt{ATMO} \citep{tremblin_cloudless_2016, tremblin_cloudless_2017} shares a similar approach for modeling the atmospheres, but it
assumes that the thermo-chemical instability of the CO/CH$_4$ and N$_2$/NH$_3$ transitions can reduce the temperature gradient in exoplanet atmospheres, acting in a similar way as clouds for reddening the spectra, but solving the issue of the J band brightening at the L/T transition \citep{burrows_l_2006}. This process is controlled by an effective adiabatic index, $\gamma$, which is lower than for the thermodynamic equilibrium. 

Both models are computed for a range of effective temperature, T$_{eff}$, surface gravity, $\mathrm{log} (g)$, metallicity, and C/O ratio. 
The details of the grid used in ours analysis are described in \citep{petrus_x-shyne_2023}. 
Since MIRI coronagraphic data only provide photometry, we intentionally compared them to near-IR photometric data from the literature and we reduced the parameter space to two: T$_{eff}$ and  $\mathrm{log} (g$), together with $\gamma$ in the case of \texttt{ATMO}, leaving the metallicity and C/O ratio for further investigations (here assuming solar values).

\begin{figure*}[ht]
    \centering
    \includegraphics[width=9cm]{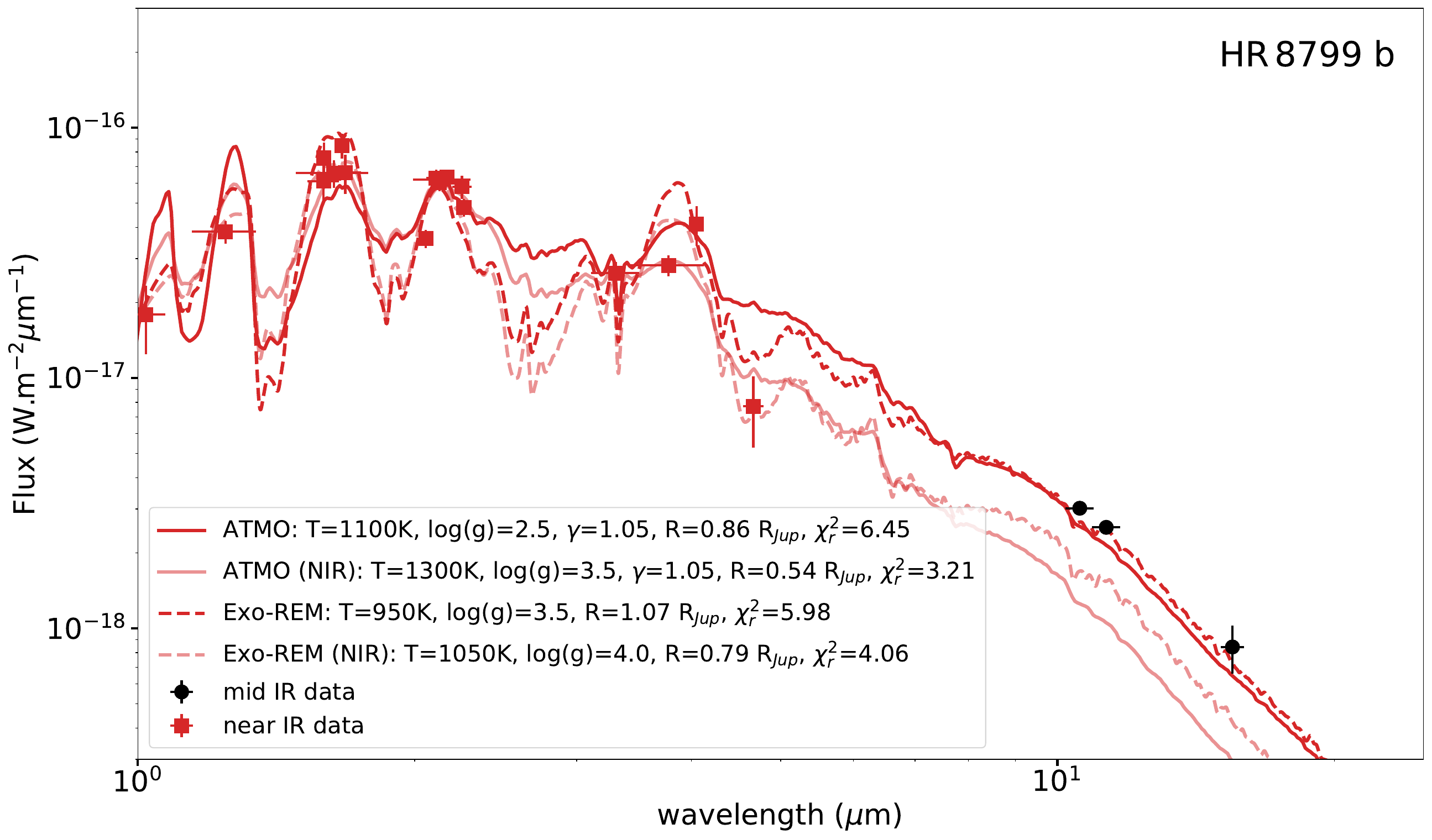}
    \includegraphics[width=9cm]{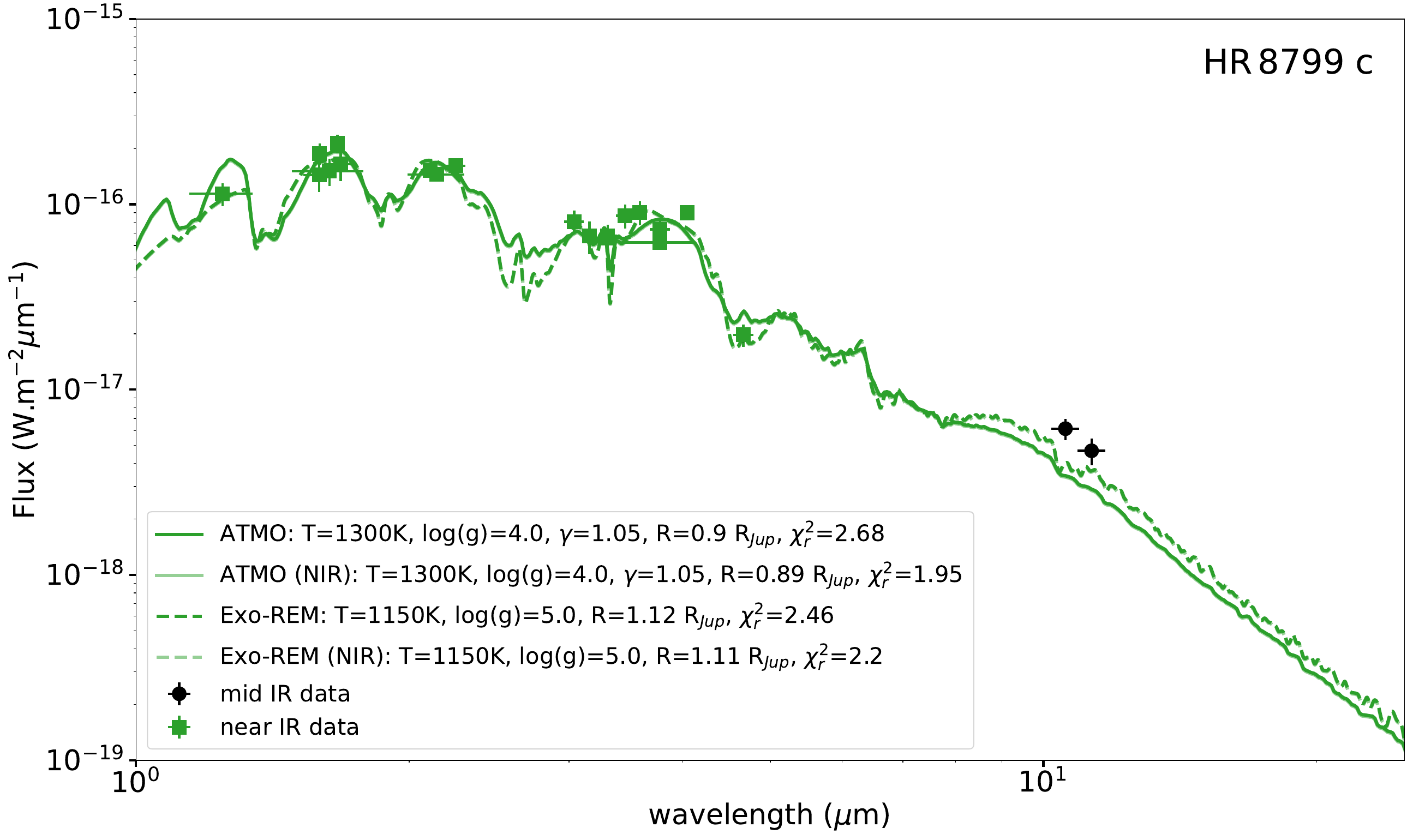}
    \includegraphics[width=9cm]{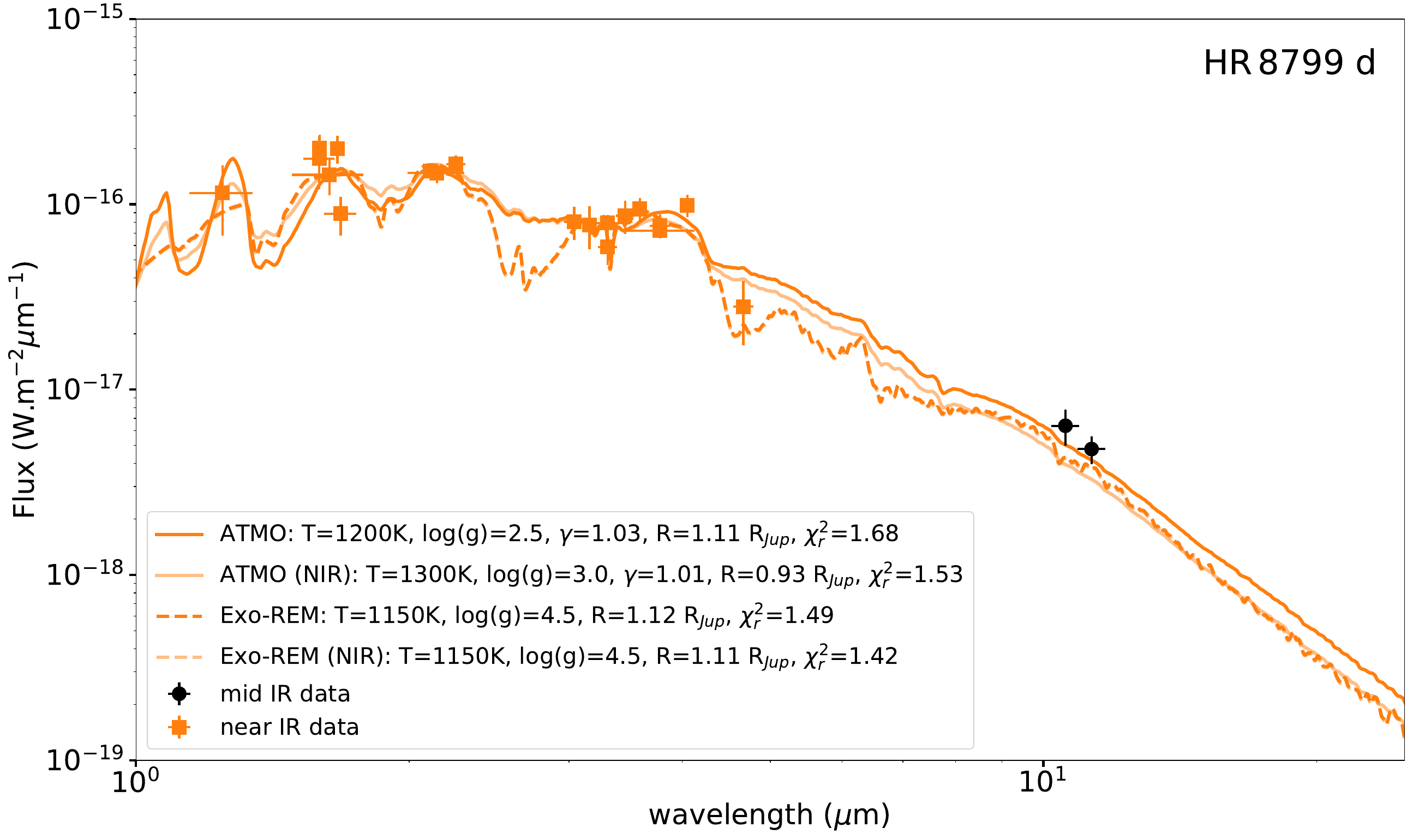}
    \includegraphics[width=9cm]{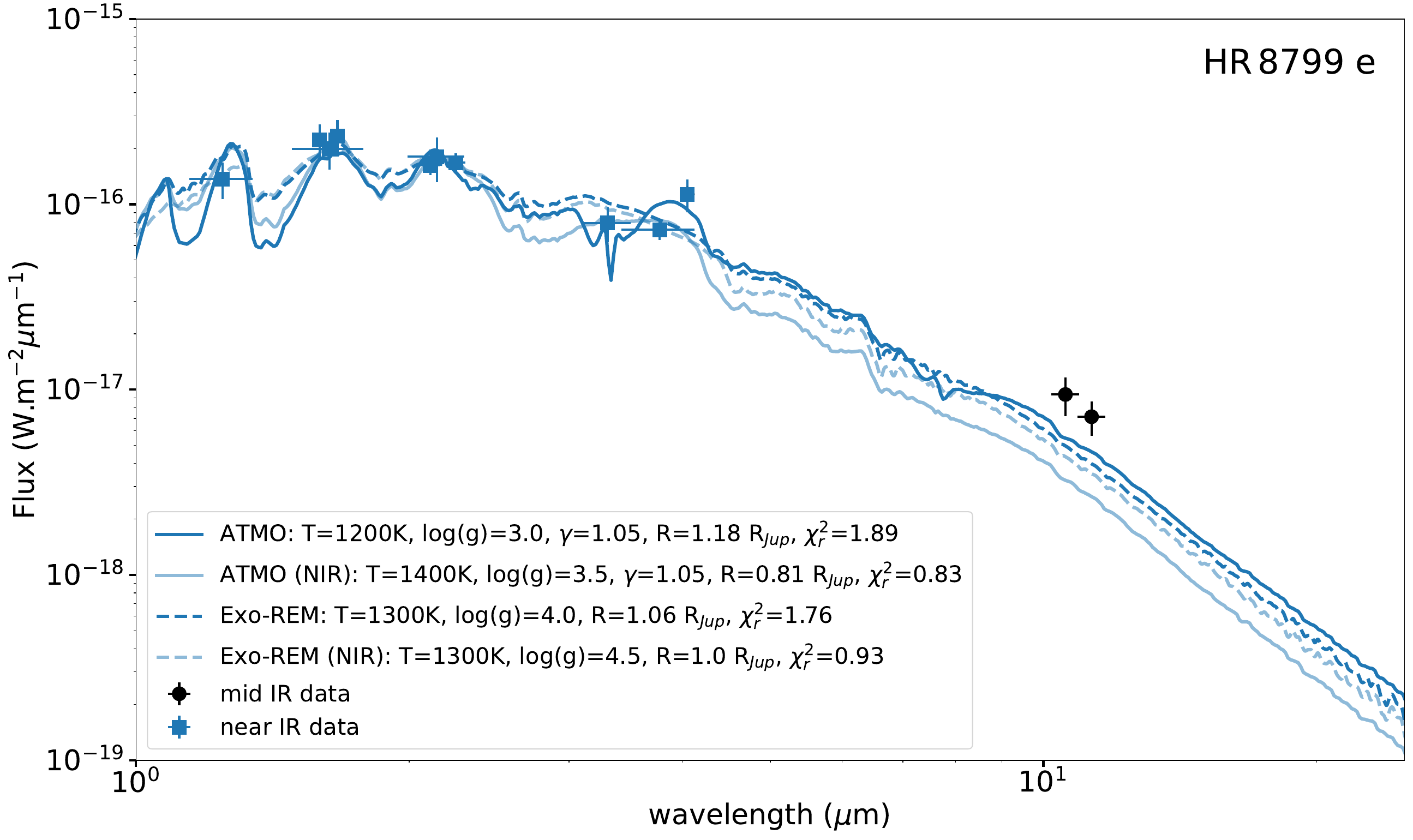}
    \caption{Flux density in W.m$^{-2}.\muup$m$^{-1}$ of the best models using \texttt{ATMO} (plain lines) and \texttt{Exo-REM} (dashed lines) fitting the near-IR (colored squares) and mid-IR (black circles) photometric data points for the four planets. The light color lines correspond to the fit of the near-IR data alone.}
    \label{fig:modelplanet}
\end{figure*}

In Figure \ref{fig:modelplanet} we compare the result of the $\chi^2$ minimization for the four planets using the two models and near-IR data alone, or using both the near-IR and mid-IR data points. As the mid-IR flux is directly proportional to the product of the effective temperature and the emitting surface of the planet, using only mid-IR photometry would obviously result in a strong degeneracy between temperature and radius. The radius determines the intensity scaling factor between the data and the model and corresponds to the value which nulls the derivative of the $\chi^2$. It reads as follows : 
\begin{equation}
R^2 =  \frac{\sum_\lambda S(\lambda) \times M(\lambda)/\sigma(\lambda)^2}{\sum_\lambda M(\lambda)^2/\sigma(\lambda)^2}
\end{equation}
with $S(\lambda)$ the data (planet's emergent spectrum), $M(\lambda)$ the model, and $\sigma(\lambda)$ the errors, evaluated in the spectral bandpasses of the photometric filters. 

Overall, the two models, \texttt{ATMO} and \texttt{Exo-REM}, qualitatively reproduce the spectral slope well, but yield large reduced $\chi^2$ values indicative of a poor fit in most cases. In general, they systematically predict lower fluxes in the $10-15\,\muup$m range than the measured photometric values. This is particularly clear for planet e. Including the mid-IR photometry in the fit together with the near-IR also provide larger reduced $\chi^2$ than near-IR alone, but this is not surprising for such a large spectral range covering more than an order of magnitude.

Planet b is the only case for which adding the mid-IR photometry makes a net difference in the fit compared to using the near-IR photometry alone. However, in this particular case, we suspect that the fit of the near-IR data could be impacted by the photometry in the M band filter (4.67\,$\muup$m) possibly explaining the low flux beyond 5\,$\muup$m. The rather low flux in this spectral band measured from the ground \citep[Keck data,][]{galicher_m-band_2011} calls for further investigations, for instance with JWST/NIRCAM data. 
The combination of MIRI and near-IR ground-based photometry provides a larger estimate of the planet b radius compared to the literature (based on near-IR data): 0.86 and 1.07\,$R_{\mathrm{J}}$, for respectively \texttt{ATMO} and \texttt{Exo-REM}, as opposed to 0.54 and 0.79\,$R_{\mathrm{J}}$. It appears more compliant with the expectation of evolutionary models \citep{phillips_new_2020}, although not yet at the predicted value for such an age ($\sim 1.3\,R_{\mathrm{J}}$). In addition, the temperature of planet b would be cooler, 1100\,K and 950\,K for \texttt{ATMO} and \texttt{Exo-REM}, respectively, instead of 1300\,K and 1050\,K. 

\begin{figure}[ht]
    \centering
    \includegraphics[width=9cm]{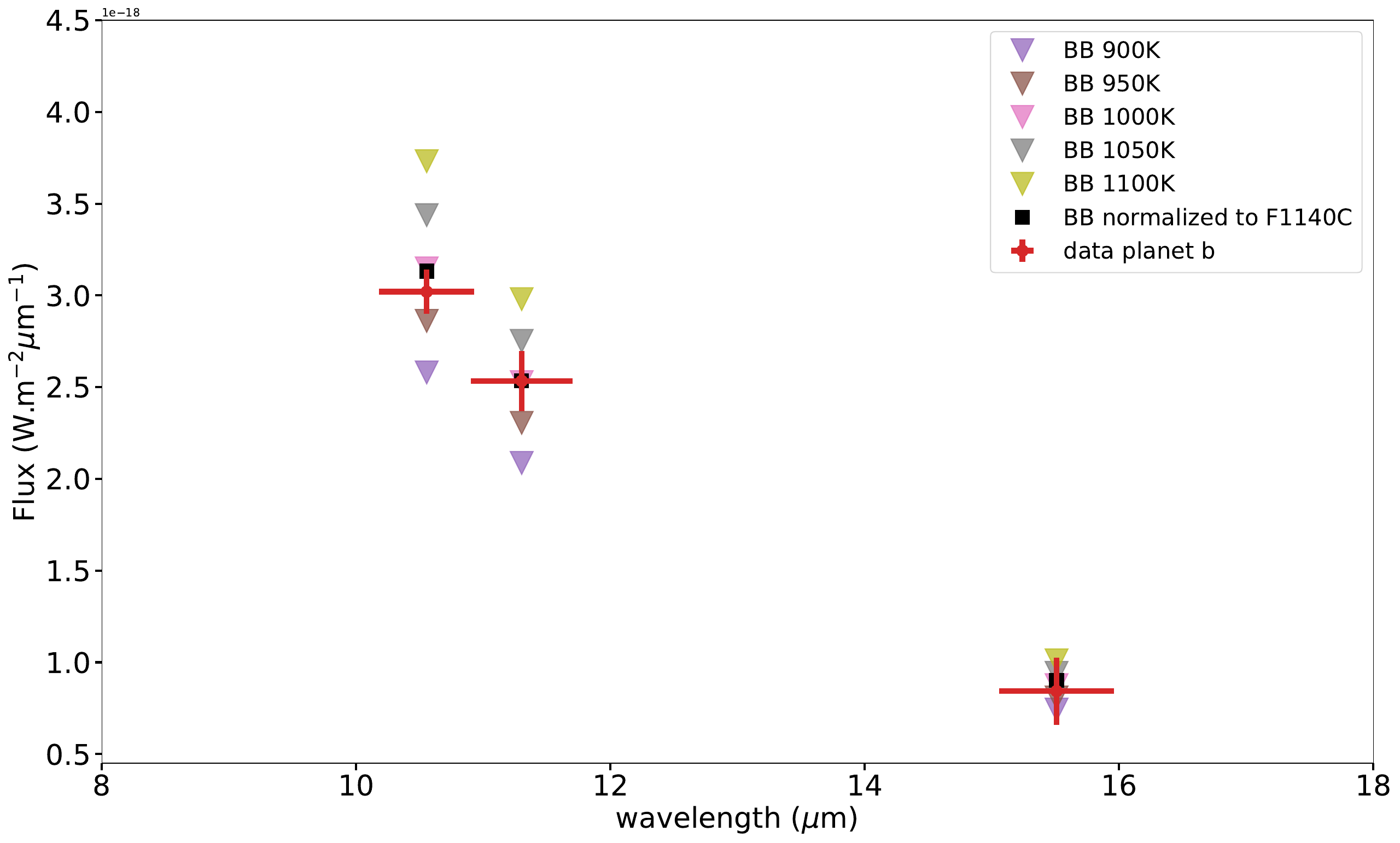}
     \caption{Flux density (W.m$^{-2}.\muup$m$^{-1}$) of HR\,8799\,b (red circles) compared to that of the blackbody for several temperatures (triangles), and if normalized to F1140C (black squares).}
    \label{fig:compare2BB}
\end{figure}

The same tendency (lower temperatures, larger radii) is also  observed, but marginally, for the other planets when incorporating mid-IR data in the fit.
We hypothesize that a possible reason for this difference in the mid-IR is because the estimation of error bars are not consistent between all the photometric data points in the literature (likely underestimated). There can be calibration issues among instruments, as well, or the models are not fully representative. In the case of HR\,8799 the more numerous near-IR measurements are naturally driving the models to converge to lower radii and higher temperatures.

For planets b and d, we note that \texttt{ATMO} produces very small and possibly un-physical surface gravity which can be as low as $\mathrm{log} (g)=2.5$, while \texttt{Exo-REM} values are more in the higher range, also with possible un-physical values as high as $\mathrm{log}(g) = 5.0$ for planet c. However, trying to derive the surface gravity with a few photometric data points is likely difficult as it mostly affects the depth of spectral features which we are not sensitive to with MIRI coronagraphy.  
Apart from planet b, the temperatures derived for the other planets are consistent with the values derived by \citet{bonnefoy_first_2016}, in the range of 1150-1300\,K.

The coronagraphic filters were originally designed to investigate the ammonia feature at $\sim10.6\,\muup$m \citep{boccaletti_mid-infrared_2015}, although the temperatures of the HR\,8799 planets are higher than the temperature at which we expect ammonia to clearly stand out in an exoplanet spectrum. Nevertheless, following \citet{danielski_atmospheric_2018} we compare the measured photometry  to blackbody spectra in order to distinguish a potential variation of the spectral slope which could be flatter at the shortest coronagraphic filter if ammonia were detectable. In Fig. \ref{fig:compare2BB}, we plot HR\,8799\,b photometry against several blackbody temperatures from 900\,K to 1100\,K assuming a radius of 1\,$R_{\mathrm{J}}$. Given the error bars, the F1065C and F1140C are the two relevant filters to derive the equivalent blackbody temperature, which would range from about 950\,K to 1000\,K. We may suspect a marginal difference at F1065C between the data and the blackbody, but, once the blackbody is normalized to the F1140C data point, the expected blackbody flux at F1065C is fully compatible with the data within error bars. The same applies to the other three planets, which have even larger photometric error bars. As a result, the current data cannot conclude on the detectability of the ammonia feature in the HR\,8799 planets.

\begin{figure*}[ht]
    \centering    \includegraphics[width=18cm]{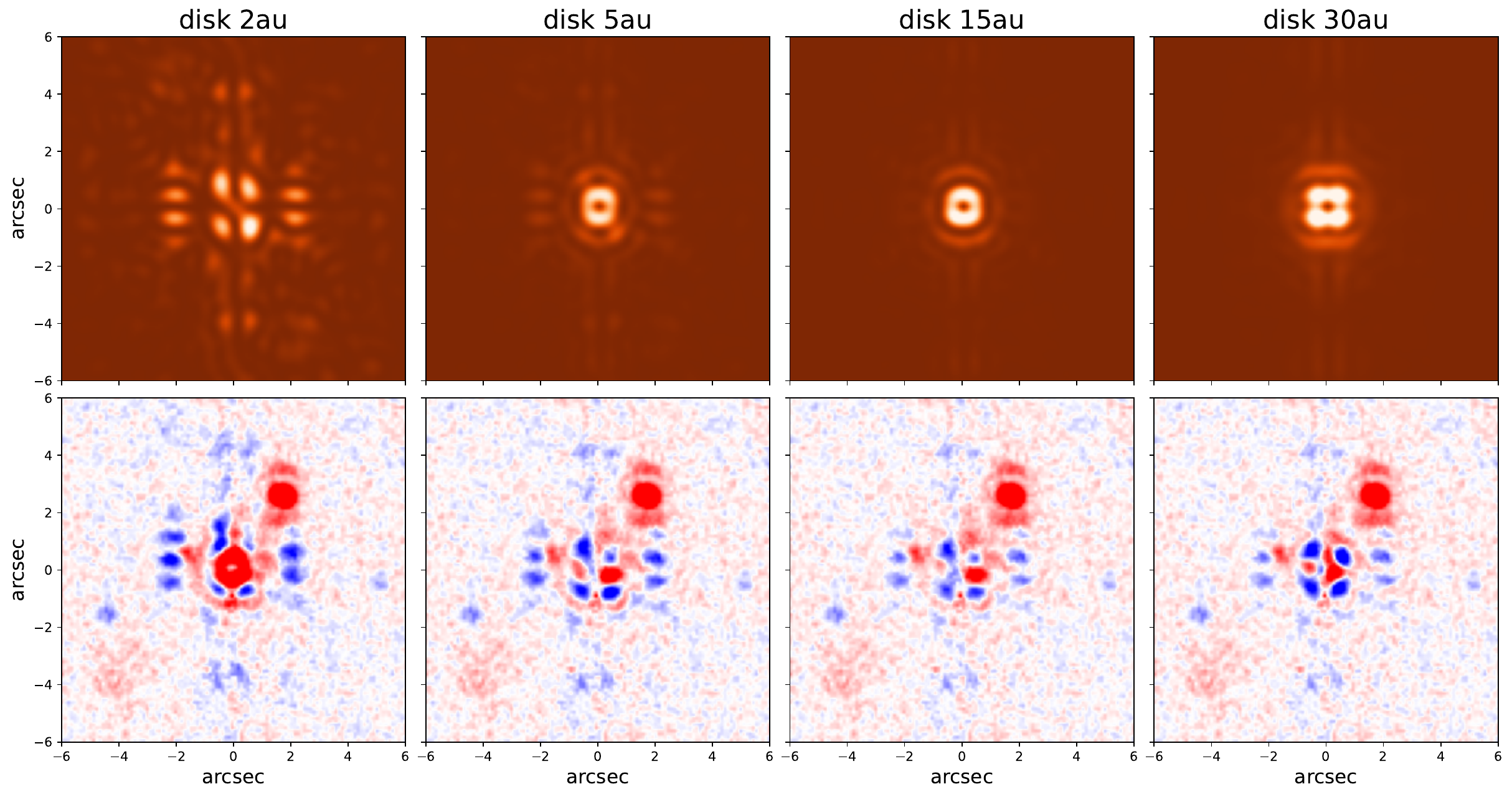}
    \caption{Coronagraphic images (top) simulated in the F1550C filter for four different disk sizes (2, 5, 15 and 30\,au), together with the residuals after subtracting the model to the real F1550C data (reference star subtracted). Intensity scale is adapted in each panel for visualisation purpose.}
    \label{fig:innerdisk}    
\end{figure*}

\section{The inner disk component}
\label{sec:disk}
At F1550C, the central part of the image is  dominated by a roughly azimuthally symmetrical pattern in the form of a broad ring surrounded by another thinner ring, which prevents the detection of the innermost planets c, d, and e. The same pattern is also visible, but fainter, in the two other coronagraphic filters, but with a reduced size. 

In fact, the warm inner disk component identified with {Spitzer} is located at physical separations of 6 to 15\,au in radius \citep{su_debris_2009} that is $0.15''$ to $0.37''$, to be compared with the angular resolution of about $0.5''$ at 15.5\,$\muup$m. In principle the inner disk should be unresolved by MIRI. 
However, such an angular size combined with the extreme sensitivity of the 4QPM coronagraph to pointing, due to its small inner working angle (which is identical to the angular resolution), results in an significant leakage of the inner disk itself (the leakage from the star being much lower in intensity). As a consequence, the inner disk image takes the form of a diffraction pattern, although with a dark spot in the central diffraction peak owing to the strong attenuation of the innermost on-axis beams.
Therefore, the rings in the image, which are more extended than the actual inner disk size, are reminiscent of a diffraction effect (scaling with wavelengths as proven by the signature at shorter wavelength), the broad ring and the dark spot being the equivalent of the central PSF diffraction peak, and the thinner ring its first diffraction ring. 
This is the same effect that hampered the detection of the inner region of the Fomalhaut disk at F1550C, as presented in \citet{gaspar_spatially_2023}.

Estimating the flux of the inner disk requires modelling the effect of the coronagraph, as a function of the disk size. As already mentioned, this disk component is angularly small, and so it is difficult to constrain its morphology with the present data. Therefore, for the sake of simplicity, we assume a uniform face-on disk model (neglecting the system's orientation), defined by a single parameter, its radius, ranging from 2 to 30\,au. We use the diffraction model, as for modeling a planet's PSF, to generate synthetic disk images. 
The 2\,au case corresponds to a perfectly unresolved case, hence the image is similar to a raw coronagraphic image of a point source (Fig. \ref{fig:innerdisk}), but at 5\,au the disk yields a strong leakage, featuring the same kind of pattern that is visible in the real data. Increasing the disk radius does not change the size of its image too much as long as it is fully dominated by diffraction effects. For larger radii ($\gtrsim$ 30\,au), we recover a more classical image of an extended source in which the 4QPM transitions generate a shadow pattern on top of the disk image. The top panels in Fig. \ref{fig:innerdisk} are to be compared with panel 3) in Fig. \ref{fig:raw_and_corono_images}. 

Minimizing the residuals between the real image and the model in a 1.65$''$ aperture radius (which encompasses the first diffraction ring of the disk), we find an optimal size of 15\,au, but 10\,au and 20\,au also provide a good match. In this case, the 4QPM attenuates the disk by a factor of 7 (5 to 13 for the extreme values), which corresponds to a total flux density of $4.2\pm0.3$\,mJy, if we assume the star flux density at 15.5\,$\muup$m to be 154.2\,mJy (interpolating between WISE:W3 and AKARI:L18W). A more realistic, ring-like disk extending from 6 to 15\,au, as in \cite{su_debris_2009}, yields similar results with a total flux density of 3.3\,mJy, and residuals that are almost identical to the uniform 15\,au case. 

The inner warm disk is also detected at F2100W. It is seen as a resolved central emission without any particular structure after subtracting the reference star (Fig. \ref{fig:raw_and_corono_images}-4). There is no coronagraphic attenuation in that case to account for, so the disk flux density can be directly integrated in an aperture of 1.65$''$. Assuming the star flux density at 21\,$\muup$m to be 101.8\,mJy (interpolating between AKARI:L18W and WISE:W4), we measure a disk flux density of $9.4$\,mJy.

To first order, these values are qualitatively in agreement with the Spitzer spectrum presented in \citet{su_debris_2009}, but a factor of about 2 lower. \citet{su_debris_2009} reported photosphere-subtracted flux densities of $\sim$8\,mJy and $\sim$19\,mJy, at respectively 15.5\,$\muup$m and 21\,$\muup$m. The exact reason of this discrepancy is still to be investigated but could be related either to the reliability of the photosphere subtraction when modeling the spectral energy distribution of the star in the presence of background objects such as redshifted galaxies (which can be confused with the star in the Spitzer beam size), to the limit of our disk model which does not capture the exact belt geometry, or to the entangling of the disk and planets' images.

\begin{figure}[ht]
    \centering
    \includegraphics[width=9cm]{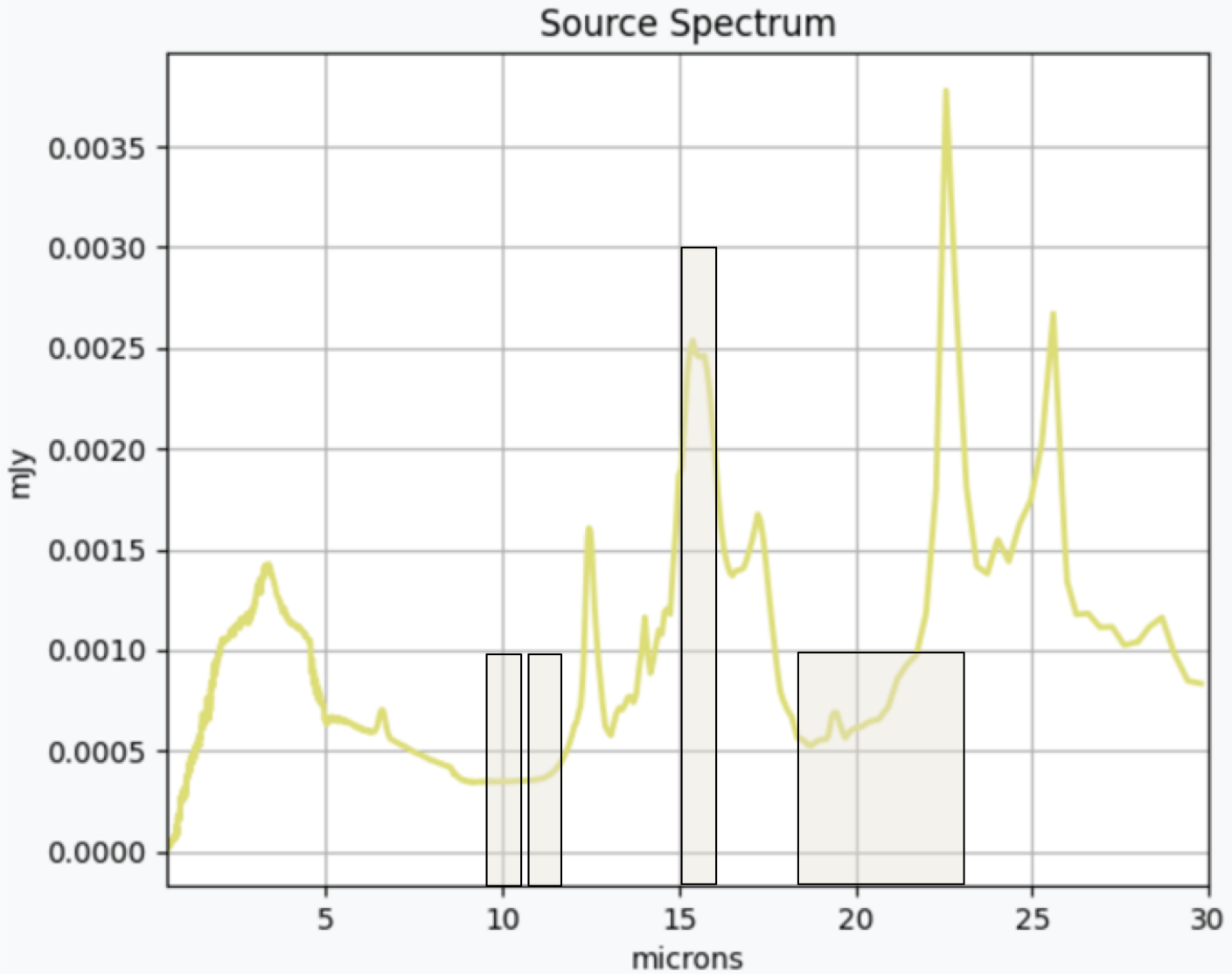}
    \caption{Typical spectrum of a Sb galaxy at a redshift z = 1, taken from the SWIRE template library (Polletta et al. 2007, ApJ, 663, 81) ; the location of the four MIRI filters are indicated. A clear  excess at $15.5 \muup$m can be observed that could explain the observed fluxes.   }
    \label{fig:spectrum_Sb_gal}
\end{figure}

\section{A background object}
\label{sec:galaxy}

Searching for additional planets in the HR\,8799 system either closer in or further out, is motivated by the structure of the debris disk made of two belts. In particular, the inner edge of the outer belt which, depending on studies, is located at a distance of 110\,au \citep{wilner_resolved_2018} or 145\,au \citep{booth_resolving_2016}, could be caused by a fifth sub-Jupiter to Saturn mass planet \citep{read_shaping_2018}. 

In this respect, the F1550C image reveals a bright point source located at $\Delta\alpha=-1.653\pm0.012''$ and $\Delta\delta=2.503\pm0.020''$ from the star (error bars being conservative since we only have two measurements), which also has fainter counterparts in the other filters. Its photometry in the four MIRI bands is shown in Fig. \ref{fig:allplanets}. With a flux density increasing from 10\,$\muup$m to 15\,$\muup$m, it could have been the signature of an object with a low temperature, possibly a planet. However, for the flux to match the expected emission of a planet cooler than the other planets in the system (400\,K for instance) it would require a  radius much larger than that of Jupiter which is odd given the mass and temperature of the other planets. 

An alternative to the planet hypothesis is provided in \cite{faramaz_detailed_2021} who identified a point-like source in ALMA band 7 data from 2018, near the inner edge of the outer belt ($\Delta\alpha=-1.28\pm0.05''$ and $\Delta\delta=2.34\pm0.05''$) with flux densities of $316\,\muup$Jy and $58\,\muup$Jy at, respectively, 0.87\,mm (band 7) and 1.3\,mm (band 6). They also showed that the spectral index is too steep to match the expected behaviour of a dust clump. As for the astrometric analysis with respect to the ALMA 2015 data (band 6) in which the same source is marginally detected, the beam size of ALMA does not allow a firm conclusion regarding the motion of the source. Interestingly, the source in the ALMA deconvolved image also seems marginally resolved. \cite{faramaz_detailed_2021} concluded that the bright source is likely a background galaxy.

With a new data point, 4.44 years apart from the former detection, we can now safely conclude that this is a background object, as the motion between the two epochs is $\Delta\alpha=-0.373\pm0.051''$ and $\Delta\delta=0.163\pm0.054''$, consistent within $\sim2\sigma$ with what is expected from the star's proper motion: $\Delta\alpha=-0.48''$ and $\Delta\delta=0.22''$. As a consequence, given that the radial separation increases from 2.67$''$ to 3$''$, it is not consistent with an orbital motion for a nearly face-on system. 

Furthermore, the characteristics of the spectral energy distribution displayed in Fig. \ref{fig:allplanets} which is peaking at $15\,\muup$m is in line with the $7.7\,\muup$m PAH emission of a Sb spiral galaxy redshifted at $z\approx 1$. As a qualitative illustration, Fig. \ref{fig:spectrum_Sb_gal} displays a typical spectrum of a Sb galaxy taken from the SWIRE template library \citep{polletta_spectral_2007}. In addition, if one compares the relative fluxes of the object in ALMA band 6 and 7 with the predictions by the redshifted Sb galaxy spectrum, normalized at $10.65\,\muup$m, they are fully consistent (ratio 1.0 and 1.12 respectively).   
In future observations with MIRI's coronagraph we should expect contamination from background galaxies to complicate the analysis.  

\section{Conclusions}
\label{sec:conclusion}
We summarize below the results of the first Guaranteed Time Observations with the MIRI coronagraphic mode. 

\begin{itemize}
    \item[$\bullet$] We obtained mid-IR observations from 10 to 20\,$\muup$m of the multiplanet system HR\,8799, in both coronagraphy and standard imaging with JWST/MIRI. 
    Once the diffraction and background emission are subtracted out we can easily detect the four planets, including the closest one in filters F1065C and F1140C. Indeed, planet e is lying at the Inner Working Angle and is attenuated by a factor of $\sim2$. This demonstrates the ability of the MIRI coronagraph to probe the very inner regions of exoplanetary systems. 

    \item[$\bullet$] We developed a procedure to extract and calibrate the photometry of the planets making use of target acquisition data and commissioning data.  
    We identified the oversubtraction as a potential source of error and mitigate this effect by masking the planets in the optimization of the reference star subtraction.
    Comparing with models of exoplanet atmospheres we show that MIRI photometry favors larger radii and cooler temperatures, than if using near-IR photometry alone. This is particularly true for planet b (0.86 or 1.07\,$R_\mathrm{J}$, and 950 or 1100\,K), but remains marginal for the other planets.  Overall, MIRI photometric values are systematically brighter than the models which could indicate a natural difficulty to compile photometries from various instruments, or a missing ingredient in the models. Taking advantage of NIRCAM to get overlapping data at near-IR, as for HIP\,65426 b \citep{carter_jwst_2023}, can definitely help to cross calibrate near and mid-IR data. 
    A deeper exploration of the atmospheric properties based on advanced modeling, taking into account these new MIRI's photometric data, would definitely be relevant.
    
    \item[$\bullet$] In terms of performance, we tested the ability to use a library of reference stars. Even if we managed to obtain reasonable contrasts, the best quality is by far achieved with a dedicated observation of a reference star obtained close in time. This situation will certainly change along the lifetime of JWST, as more targets are observed with MIRI’s coronagraphs, providing sufficient observing time is allocated to this mode. 

    \item[$\bullet$] In line with the capacity of the MIRI’s coronagraph at short angular separations, we also directly detect and spatially resolve for the first time the inner warm debris disk in all four filters. By taking into account the diffractive effect of the 4QPM coronagraph we were able to constrain the inner disk radius to about $15\pm5$\,au using the F1550C image, which provides a direct and independent estimate in agreement with the modeling of the IR excess. We measured a flux density at F1550C and F2100W of a few mJy, a factor of 2 lower than { Spitzer} which remains to be understood. The outer cold debris disk is undetected, consistent with the sensitivity.

    \item[$\bullet$] In the field of view, we identified a point source which we confirmed to be a background redshifted galaxy at $z\approx1$ considering the astrometry with respect to ALMA observations, and its spectral energy distribution. 

    \item[$\bullet$] Finally, the extreme sensitivity of the 4QPM coronagraph at small angular separations combined with the presence of inner circumstellar components can make
    the detection and the interpretation of young system observations very challenging, not mentioning the confusion related to background galaxies. The MIRI’s coronagraphic mode is still in its infancy, and there is room for improving several aspects: modeling and calibrating the diffraction pattern as a function of telescope characteristics, developing optimal data reduction techniques, and interpreting the entangled  signals of point-like sources and extended circumstellar components. 

\end{itemize}


\begin{appendix}

\section {Other data reductions}
\label{app:other_red}

\begin{figure*}
    \centering
    \includegraphics[height=22cm]{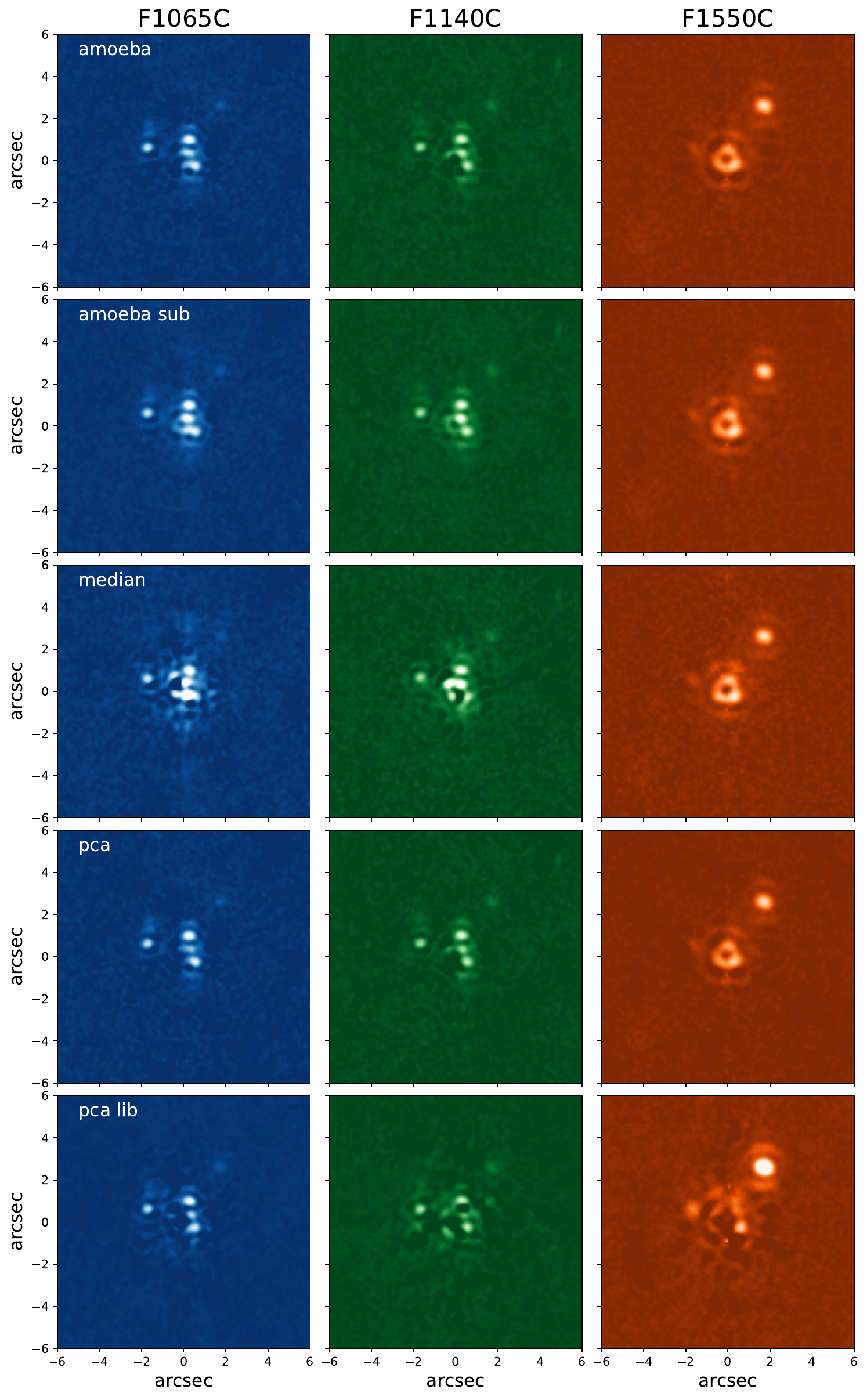}
    \caption{
Reference star subtracted images in the three coronagraphic filters (left to right: F1065C, F1140C, F1550C) for a variety of algorithms: amoeba (same as in Fig. \ref{fig:raw_and_corono_images} without masking the planets), amoeba sub. (same as in Fig. \ref{fig:raw_and_corono_images} with subtracting the planets' diffraction models, median (median combination of the 9 small grid dithers), PCA (principal component analysis of the 9 small grid dithers), PCA lib (principal component analysis using two other reference stars observed at commissioning, hence totalling 18 small grid dithers). The Field of View is $12\times 12"$. Intensity scale is adapted in each panel for visualisation purpose.}
    \label{fig:allalgo}
\end{figure*}

\section {Planets' footprints}
\label{app:other_red}
\begin{figure*}
    \centering
    \includegraphics[width=18cm]{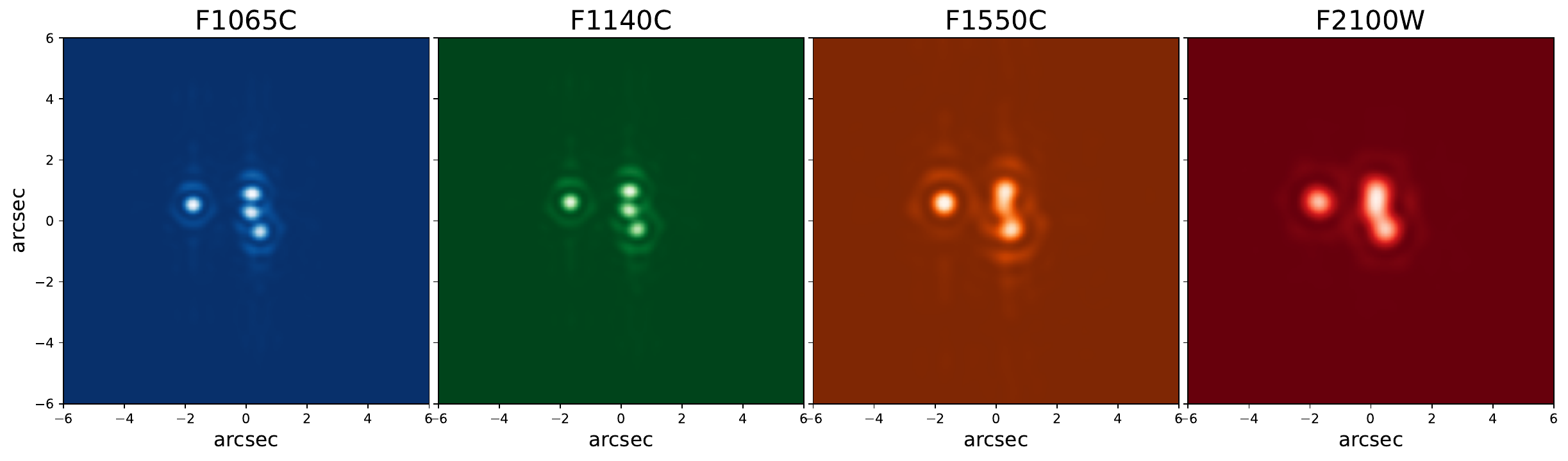}
    \caption{Images in each filters of four point sources with equal brightness, but located at the positions of the HR8799 planets to illustrate the impact of the coronagraph transmission as well as the overlapping of the PSFs arising from the angular resolution at such angular separations. }
    \label{fig:planetsalone}
\end{figure*}

\section {Measured contrasts}

\begin{table*}[]
    \centering
    \begin{tabular}{lllll}
        planet                      & norm.              & F1065C           & F1140C           & F1550C  \\
                                    & \& extrac. method  &                  &                  &   \\ \hline \hline
        \multirow{ 4}{*}{b}         & CORO / aper        &  $2608\pm101$    &  $2491\pm71$     &    $1782\pm50$      \\           
                                    & CORO / FP neg      &  $2618\pm102$    &  $2317\pm66$     &    $2650\pm75$      \\ 
                                    & TA / aper          &  $2471\pm20$     &  $2113\pm19$     &    $1221\pm16$      \\ 
                                    & TA / FP neg        &  $2480\pm20$     &  $1965\pm17$     &    $1817\pm24$      \\ \hline
        \multirow{ 4}{*}{c}         & CORO / aper        &  $1142\pm44$     &  $1133\pm32$     &    $>415$           \\            
                                    & CORO / FP neg      &  $1466\pm57$     &  $1532\pm44$     &    $>589$           \\ 
                                    & TA / aper          &  $1082\pm9$      &  $961\pm8$       &    $>284$           \\ 
                                    & TA / FP neg        &  $1389\pm11$     &  $1299\pm11$     &    $>404$           \\ \hline
        \multirow{ 4}{*}{d}         & CORO / aper        &  $1025\pm40$     & $1101\pm31$      &    $>328$           \\            
                                    & CORO / FP neg      &  $1503\pm58$     &  $1511\pm43$     &    $>467$           \\ 
                                    & TA / aper          &  $971\pm8$       &  $934\pm8$       &    $>225$           \\ 
                                    & TA / FP neg        &  $1424\pm12$     & $1282\pm11$      &    $>320$           \\ \hline
        \multirow{ 4}{*}{e}         & CORO / aper        &  $685\pm27$      &  $712\pm2$      &    $>164$           \\            
                                    & CORO / FP neg      &  $1094\pm42$     &   $1072\pm31$    &    $>336$           \\ 
                                    & TA / aper          &  $649\pm5$       &  $604\pm5$       &    $>112$           \\ 
                                    & TA / FP neg        &  $1037\pm8$      &  $909\pm8$       &    $>230$           \\  \hline \hline
               
    \end{tabular}
    \caption{Same as Table \ref{tab:contrasts_mean} for each normalization/extraction method.}
    \label{tab:contrasts}
\end{table*}

\end{appendix}

\begin{acknowledgement}
This work is based on observations made with the NASA/ESA/CSA James Webb Space Telescope. The data were obtained from the Mikulski Archive for Space Telescopes at the Space Telescope Science Institute, which is operated by the Association of Universities for Research in Astronomy, Inc., under NASA contract NAS 5-03127 for JWST. These observations are associated with program \#1194.
Part of this work was carried out at the Jet Propulsion Laboratory, California Institute of Technology, under contrast with NASA (80NM0018D0004).
French contributors acknowledge the support and funding of CNES.
This publication makes use of VOSA, developed under the Spanish Virtual Observatory (https://svo.cab.inta-csic.es) project funded by MCIN/AEI/10.13039/501100011033/ through grant PID2020-112949GB-I00. 
VOSA has been partially updated by using funding from the European Union's Horizon 2020 Research and Innovation Programme, under Grant Agreement nº 776403 (EXOPLANETS-A)
This work has made use of data from the European Space Agency (ESA) mission
{\it Gaia} (\url{https://www.cosmos.esa.int/gaia}), processed by the {\it Gaia}
Data Processing and Analysis Consortium (DPAC,
\url{https://www.cosmos.esa.int/web/gaia/dpac/consortium}). Funding for the DPAC
has been provided by national institutions, in particular the institutions
participating in the {\it Gaia} Multilateral Agreement.
This research has made use of the VizieR catalogue access tool, CDS,
 Strasbourg, France (DOI : 10.26093/cds/vizier). The original description 
 of the VizieR service was published in 2000, A\&AS 143, 23

\end{acknowledgement}

\bibliographystyle{aa}
\bibliography{mylib}

\begin{thebibliography}{62}
\expandafter\ifx\csname natexlab\endcsname\relax\def\natexlab#1{#1}\fi

\bibitem[{Allard {et~al.}(2011)Allard, Homeier, \& Freytag}]{allard_model_2011}
Allard, F., Homeier, D., \& Freytag, B. 2011, in Astronomical {Society} of the
  {Pacific} {Conference} {Series}, Vol. 448, 91

\bibitem[{Asplund {et~al.}(2009)Asplund, Grevesse, Sauval, \&
  Scott}]{asplund_chemical_2009}
Asplund, M., Grevesse, N., Sauval, A.~J., \& Scott, P. 2009, Annual Review of
  Astronomy and Astrophysics, 47, 481

\bibitem[{Barman {et~al.}(2015)Barman, Konopacky, Macintosh, \&
  Marois}]{barman_simultaneous_2015}
Barman, T.~S., Konopacky, Q.~M., Macintosh, B., \& Marois, C. 2015, The
  Astrophysical Journal, 804, 61

\bibitem[{Barman {et~al.}(2011)Barman, Macintosh, Konopacky, \&
  Marois}]{barman_clouds_2011}
Barman, T.~S., Macintosh, B., Konopacky, Q.~M., \& Marois, C. 2011, The
  Astrophysical Journal, 733, 65

\bibitem[{Baudino {et~al.}(2015)Baudino, Bézard, Boccaletti, Bonnefoy,
  Lagrange, \& Galicher}]{baudino_interpreting_2015}
Baudino, J.~L., Bézard, B., Boccaletti, A., {et~al.} 2015, Astronomy \&
  Astrophysics, 582, A83

\bibitem[{Bayo {et~al.}(2008)Bayo, Rodrigo, Barrado Y~Navascués, Solano,
  Gutiérrez, Morales-Calderón, \& Allard}]{bayo_vosa_2008}
Bayo, A., Rodrigo, C., Barrado Y~Navascués, D., {et~al.} 2008, Astronomy \&
  Astrophysics, 492, 277

\bibitem[{Boccaletti {et~al.}(2005)Boccaletti, Baudoz, Baudrand, Reess, \&
  Rouan}]{boccaletti_imaging_2005}
Boccaletti, A., Baudoz, P., Baudrand, J., Reess, J.~M., \& Rouan, D. 2005,
  Advances in Space Research, 36, 1099

\bibitem[{Boccaletti {et~al.}(2022)Boccaletti, Cossou, Baudoz, Lagage, Dicken,
  Glasse, Hines, Aguilar, Detre, Nickson, Noriega-Crespo, Gáspár, Labiano,
  Stark, Rouan, Reess, Wright, Rieke, Garcia~Marin, Lajoie, Girard, Perrin,
  Soummer, \& Pueyo}]{boccaletti_jwstmiri_2022}
Boccaletti, A., Cossou, C., Baudoz, P., {et~al.} 2022, Astronomy \&
  Astrophysics, 667, A165

\bibitem[{Boccaletti {et~al.}(2015)Boccaletti, Lagage, Baudoz, Beichman,
  Bouchet, Cavarroc, Dubreuil, Glasse, Glauser, Hines, Lajoie, Lebreton,
  Perrin, Pueyo, Reess, Rieke, Ronayette, Rouan, Soummer, \&
  Wright}]{boccaletti_mid-infrared_2015}
Boccaletti, A., Lagage, P.-O., Baudoz, P., {et~al.} 2015, Publications of the
  Astronomical Society of the Pacific, 127, 633

\bibitem[{Bonnefoy {et~al.}(2016)Bonnefoy, Zurlo, Baudino, Lucas, Mesa, Maire,
  Vigan, Galicher, Homeier, Marocco, Gratton, Chauvin, Allard, Desidera,
  Kasper, Moutou, Lagrange, Antichi, Baruffolo, Baudrand, Beuzit, Boccaletti,
  Cantalloube, Carbillet, Charton, Claudi, Costille, Dohlen, Dominik, Fantinel,
  Feautrier, Feldt, Fusco, Gigan, Girard, Gluck, Gry, Henning, Janson,
  Langlois, Madec, Magnard, Maurel, Mawet, Meyer, Milli, Moeller-Nilsson,
  Mouillet, Pavlov, Perret, Pujet, Quanz, Rochat, Rousset, Roux, Salasnich,
  Salter, Sauvage, Schmid, Sevin, Soenke, Stadler, Turatto, Udry, Vakili,
  Wahhaj, \& Wildi}]{bonnefoy_first_2016}
Bonnefoy, M., Zurlo, A., Baudino, J.~L., {et~al.} 2016, Astronomy \&
  Astrophysics, 587, A58

\bibitem[{Booth {et~al.}(2016)Booth, Jordán, Casassus, Hales, Dent, Faramaz,
  Matrà, Barkats, Brahm, \& Cuadra}]{booth_resolving_2016}
Booth, M., Jordán, A., Casassus, S., {et~al.} 2016, Monthly Notices of the
  Royal Astronomical Society: Letters, 460, L10

\bibitem[{Burrows {et~al.}(2006)Burrows, Sudarsky, \& Hubeny}]{burrows_l_2006}
Burrows, A., Sudarsky, D., \& Hubeny, I. 2006, The Astrophysical Journal, 640,
  1063

\bibitem[{Carter {et~al.}(2023)Carter, Hinkley, Kammerer, Skemer, Biller,
  Leisenring, Millar-Blanchaer, Petrus, Stone, Ward-Duong, Wang, Girard, Hines,
  Perrin, Pueyo, Balmer, Bonavita, Bonnefoy, Chauvin, Choquet, Christiaens,
  Danielski, Kennedy, Matthews, Miles, Patapis, Ray, Rickman, Sallum,
  Stapelfeldt, Whiteford, Zhou, Absil, Boccaletti, Booth, Bowler, Chen, Currie,
  Fortney, Grady, Greenbaum, Henning, Hoch, Janson, Kalas, Kenworthy, Kervella,
  Kraus, Lagage, Liu, Macintosh, Marino, Marley, Marois, Matthews, Mawet,
  McElwain, Metchev, Meyer, Molliere, Moran, Morley, Mukherjee, Pantin,
  Quirrenbach, Rebollido, Ren, Schneider, Vasist, Worthen, Wyatt,
  Briesemeister, Bryan, Calissendorff, Cantalloube, Cugno, De~Furio, Dupuy,
  Factor, Faherty, Fitzgerald, Franson, Gonzales, Hood, Howe, Kuzuhara,
  Lagrange, Lawson, Lazzoni, Lew, Liu, Llop-Sayson, Lloyd, Martinez, Mazoyer,
  Palma-Bifani, Quanz, Redai, Samland, Schlieder, Tamura, Tan, Uyama, Vigan,
  Vos, Wagner, Wolff, Ygouf, Zhang, Zhang, \& Zhang}]{carter_jwst_2023}
Carter, A.~L., Hinkley, S., Kammerer, J., {et~al.} 2023, The {JWST} {Early}
  {Release} {Science} {Program} for {Direct} {Observations} of {Exoplanetary}
  {Systems} {I}: {High} {Contrast} {Imaging} of the {Exoplanet} {HIP} 65426 b
  from 2-16 \${\textbackslash}mu\$m, arXiv:2208.14990 [astro-ph]

\bibitem[{{Cavarroc} {et~al.}(2008){Cavarroc}, {Amiaux}, {Baudoz},
  {Boccaletti}, {Bouchet}, {Dubreuil}, {Lagage}, {Moreau}, {Pantin}, {Reess},
  {Ronayette}, \& {Wright}}]{cavarroc_first_2008}
{Cavarroc}, C., {Amiaux}, J., {Baudoz}, P., {et~al.} 2008, in Society of
  Photo-Optical Instrumentation Engineers (SPIE) Conference Series, Vol. 7010,
  Space Telescopes and Instrumentation 2008: Optical, Infrared, and Millimeter,
  ed. J.~{Oschmann}, Jacobus~M., M.~W.~M. {de Graauw}, \& H.~A. {MacEwen},
  70100W

\bibitem[{Cavarroc {et~al.}(2008)Cavarroc, Boccaletti, Baudoz, Amiaux, \&
  Regan}]{cavarroc_target_2008}
Cavarroc, C., Boccaletti, A., Baudoz, P., Amiaux, J., \& Regan, M. 2008, The
  Publications of the Astronomical Society of the Pacific, 120, 1016

\bibitem[{Charnay {et~al.}(2018)Charnay, Bézard, Baudino, Bonnefoy,
  Boccaletti, \& Galicher}]{charnay_self-consistent_2018}
Charnay, B., Bézard, B., Baudino, J.~L., {et~al.} 2018, The Astrophysical
  Journal, 854, 0

\bibitem[{Chauvin {et~al.}(2017)Chauvin, Desidera, Lagrange, Vigan, Gratton,
  Langlois, Bonnefoy, Beuzit, Feldt, Mouillet, Meyer, Cheetham, Biller,
  Boccaletti, D'Orazi, Galicher, Hagelberg, Maire, Mesa, Olofsson, Samland,
  Schmidt, Sissa, Bonavita, Charnay, Cudel, Daemgen, Delorme, Janin-Potiron,
  Janson, Keppler, Coroller, Ligi, Marleau, Messina, Mollière, Mordasini,
  Müller, Peretti, Perrot, Rodet, Rouan, Zurlo, Dominik, Henning, Menard,
  Schmid, Turatto, Udry, Vakili, Abe, Antichi, Baruffolo, Baudoz, Baudrand,
  Blanchard, Bazzon, Buey, Carbillet, Carle, Charton, Cascone, Claudi,
  Costille, Deboulbé, Caprio, Dohlen, Fantinel, Feautrier, Fusco, Gigan, Giro,
  Gisler, Gluck, Hubin, Hugot, Jaquet, Kasper, Madec, Magnard, Martinez,
  Maurel, Mignant, Möller-Nilsson, Llored, Moulin, Origne, Pavlov, Perret,
  Petit, Pragt, Puget, Rabou, Ramos, Rigal, Rochat, Roelfsema, Rousset, Roux,
  Salasnich, Sauvage, Sevin, Soenke, Stadler, Suarez, Weber, Wildi, Antoniucci,
  Augereau, Baudino, Brandner, Engler, Girard, Gry, Kral, Kopytova, Lagadec,
  Milli, Moutou, Schlieder, Szulágyi, Thalmann, \&
  Wahhaj}]{chauvin_discovery_2017}
Chauvin, G., Desidera, S., Lagrange, A.~M., {et~al.} 2017, Astronomy \&
  Astrophysics, 605, L9

\bibitem[{Chen {et~al.}(2006)Chen, Sargent, Bohac, Kim, Leibensperger, Jura,
  Najita, Forrest, Watson, Sloan, \& Keller}]{chen_spitzer_2006}
Chen, C.~H., Sargent, B.~A., Bohac, C., {et~al.} 2006, The Astrophysical
  Journal Supplement Series, 166, 351

\bibitem[{Currie {et~al.}(2011)Currie, Burrows, Itoh, Matsumura, Fukagawa,
  Apai, Madhusudhan, Hinz, Rodigas, Kasper, Pyo, \&
  Ogino}]{currie_combined_2011}
Currie, T., Burrows, A., Itoh, Y., {et~al.} 2011, The Astrophysical Journal,
  729, 128

\bibitem[{Danielski {et~al.}(2018)Danielski, Baudino, Lagage, Boccaletti,
  Gastaud, Coulais, \& Bézard}]{danielski_atmospheric_2018}
Danielski, C., Baudino, J.-L., Lagage, P.-O., {et~al.} 2018, The Astronomical
  Journal, 156, 276

\bibitem[{Faramaz {et~al.}(2021)Faramaz, Marino, Booth, Matrà, Mamajek,
  Bryden, Stapelfeldt, Casassus, Cuadra, Hales, \&
  Zurlo}]{faramaz_detailed_2021}
Faramaz, V., Marino, S., Booth, M., {et~al.} 2021, The Astronomical Journal,
  161, 271

\bibitem[{{Gaia Collaboration} {et~al.}(2021){Gaia Collaboration}, Brown,
  Vallenari, Prusti, De~Bruijne, Babusiaux, Biermann, Creevey, Evans, Eyer,
  Hutton, Jansen, Jordi, Klioner, Lammers, Lindegren, Luri, Mignard, Panem,
  Pourbaix, Randich, Sartoretti, Soubiran, Walton, Arenou, Bailer-Jones,
  Bastian, Cropper, Drimmel, Katz, Lattanzi, Van~Leeuwen, Bakker, Cacciari,
  Castañeda, De~Angeli, Ducourant, Fabricius, Fouesneau, Frémat, Guerra,
  Guerrier, Guiraud, Jean-Antoine~Piccolo, Masana, Messineo, Mowlavi, Nicolas,
  Nienartowicz, Pailler, Panuzzo, Riclet, Roux, Seabroke, Sordo, Tanga,
  Thévenin, Gracia-Abril, Portell, Teyssier, Altmann, Andrae, Bellas-Velidis,
  Benson, Berthier, Blomme, Brugaletta, Burgess, Busso, Carry, Cellino, Cheek,
  Clementini, Damerdji, Davidson, Delchambre, Dell’Oro,
  Fernández-Hernández, Galluccio, García-Lario, Garcia-Reinaldos,
  González-Núñez, Gosset, Haigron, Halbwachs, Hambly, Harrison,
  Hatzidimitriou, Heiter, Hernández, Hestroffer, Hodgkin, Holl, Janßen,
  Jevardat De~Fombelle, Jordan, Krone-Martins, Lanzafame, Löffler, Lorca,
  Manteiga, Marchal, Marrese, Moitinho, Mora, Muinonen, Osborne, Pancino,
  Pauwels, Petit, Recio-Blanco, Richards, Riello, Rimoldini, Robin, Roegiers,
  Rybizki, Sarro, Siopis, Smith, Sozzetti, Ulla, Utrilla, Van~Leeuwen,
  Van~Reeven, Abbas, Abreu~Aramburu, Accart, Aerts, Aguado, Ajaj, Altavilla,
  Álvarez, Álvarez Cid-Fuentes, Alves, Anderson, Anglada~Varela, Antoja,
  Audard, Baines, Baker, Balaguer-Núñez, Balbinot, Balog, Barache, Barbato,
  Barros, Barstow, Bartolomé, Bassilana, Bauchet, Baudesson-Stella, Becciani,
  Bellazzini, Bernet, Bertone, Bianchi, Blanco-Cuaresma, Boch, Bombrun,
  Bossini, Bouquillon, Bragaglia, Bramante, Breedt, Bressan, Brouillet,
  Bucciarelli, Burlacu, Busonero, Butkevich, Buzzi, Caffau, Cancelliere,
  Cánovas, Cantat-Gaudin, Carballo, Carlucci, Carnerero, Carrasco,
  Casamiquela, Castellani, Castro-Ginard, Castro~Sampol, Chaoul, Charlot,
  Chemin, Chiavassa, Cioni, Comoretto, Cooper, Cornez, Cowell, Crifo, Crosta,
  Crowley, Dafonte, Dapergolas, David, David, De~Laverny, De~Luise, De~March,
  De~Ridder, De~Souza, De~Teodoro, De~Torres, Del~Peloso, Del~Pozo, Delbo,
  Delgado, Delgado, Delisle, Di~Matteo, Diakite, Diener, Distefano, Dolding,
  Eappachen, Edvardsson, Enke, Esquej, Fabre, Fabrizio, Faigler, Fedorets,
  Fernique, Fienga, Figueras, Fouron, Fragkoudi, Fraile, Franke, Gai, Garabato,
  Garcia-Gutierrez, García-Torres, Garofalo, Gavras, Gerlach, Geyer, Giacobbe,
  Gilmore, Girona, Giuffrida, Gomel, Gomez, Gonzalez-Santamaria,
  González-Vidal, Granvik, Gutiérrez-Sánchez, Guy, Hauser, Haywood, Helmi,
  Hidalgo, Hilger, Hładczuk, Hobbs, Holland, Huckle, Jasniewicz, Jonker,
  Juaristi~Campillo, Julbe, Karbevska, Kervella, Khanna, Kochoska, Kontizas,
  Kordopatis, Korn, Kostrzewa-Rutkowska, Kruszyńska, Lambert, Lanza, Lasne,
  Le~Campion, Le~Fustec, Lebreton, Lebzelter, Leccia, Leclerc, Lecoeur-Taibi,
  Liao, Licata, Lindstrøm, Lister, Livanou, Lobel, Madrero~Pardo, Managau,
  Mann, Marchant, Marconi, Marcos~Santos, Marinoni, Marocco, Marshall,
  Martin~Polo, Martín-Fleitas, Masip, Massari, Mastrobuono-Battisti, Mazeh,
  McMillan, Messina, Michalik, Millar, Mints, Molina, Molinaro, Molnár,
  Montegriffo, Mor, Morbidelli, Morel, Morris, Mulone, Munoz, Muraveva, Murphy,
  Musella, Noval, Ordénovic, Orrù, Osinde, Pagani, Pagano, Palaversa,
  Palicio, Panahi, Pawlak, Peñalosa~Esteller, Penttilä, Piersimoni, Pineau,
  Plachy, Plum, Poggio, Poretti, Poujoulet, Prša, Pulone, Racero, Ragaini,
  Rainer, Raiteri, Rambaux, Ramos, Ramos-Lerate, Re~Fiorentin, Regibo, Reylé,
  Ripepi, Riva, Rixon, Robichon, Robin, Roelens, Rohrbasser, Romero-Gómez,
  Rowell, Royer, Rybicki, Sadowski, Sagristà~Sellés, Sahlmann, Salgado,
  Salguero, Samaras, Sanchez~Gimenez, Sanna, Santoveña, Sarasso, Schultheis,
  Sciacca, Segol, Segovia, Ségransan, Semeux, Shahaf, Siddiqui, Siebert,
  Siltala, Slezak, Smart, Solano, Solitro, Souami, Souchay, Spagna, Spoto,
  Steele, Steidelmüller, Stephenson, Süveges, Szabados, Szegedi-Elek, Taris,
  Tauran, Taylor, Teixeira, Thuillot, Tonello, Torra, Torra, Turon, Unger,
  Vaillant, Van~Dillen, Vanel, Vecchiato, Viala, Vicente, Voutsinas, Weiler,
  Wevers, Wyrzykowski, Yoldas, Yvard, Zhao, Zorec, Zucker, Zurbach, \&
  Zwitter}]{gaia_collaboration_gaia_2021}
{Gaia Collaboration}, Brown, A. G.~A., Vallenari, A., {et~al.} 2021, Astronomy
  \& Astrophysics, 649, A1

\bibitem[{Galicher {et~al.}(2011)Galicher, Marois, Macintosh, Barman, \&
  Konopacky}]{galicher_m-band_2011}
Galicher, R., Marois, C., Macintosh, B., Barman, T., \& Konopacky, Q. 2011, The
  Astrophysical Journal Letters, 739, L41

\bibitem[{Gardner {et~al.}(2023)Gardner, Mather, Abbott, Abell, Abernathy,
  Abney, Abraham, Abraham, Abul-Huda, Acton, Adams, Adams, Adler, Adriaensen,
  Aguilar, Ahmed, Ahmed, Ahmed, Albat, Albert, Alberts, Aldridge, Allen, Allen,
  Altenburg, Altunc, Alvarez, Álvarez Márquez, de~Oliveira, Ambrose,
  Anandakrishnan, Andersen, Anderson, Anderson, Anderson, Anderson, Aprea,
  Archer, Arenberg, Argyriou, Arribas, Artigau, Arvai, Atcheson, Atkinson,
  Averbukh, Aymergen, Bacinski, Baggett, Bagnasco, Baker, Balzano, Banks,
  Baran, Barker, Barrett, Barringer, Barto, Bast, Baudoz, Baum, Beatty,
  Beaulieu, Bechtold, Beck, Beddard, Beichman, Bellagama, Bely, Berger,
  Bergeron, Darveau-Bernier, Bertch, Beskow, Betz, Biagetti, Birkmann,
  Bjorklund, Blackwood, Blazek, Blossfeld, Bluth, Boccaletti, Boegner~Jr.,
  Bohlin, Boia, Böker, Bonaventura, Bond, Bosley, Boucarut, Bouchet, Bouwman,
  Bower, Bowers, Bowers, Boyce, Boyer, Boyer, Boyer, Boyer, Bradley, Brady,
  Brandl, Brannen, Breda, Bremmer, Brennan, Bresnahan, Bright, Broiles,
  Bromenschenkel, Brooks, Brooks, Brown, Brown, Brown, Bruce, Bryson, Bujanda,
  Bullock, Bunker, Bureo, Burt, Bush, Bushouse, Bussman, Cabaud, Cale, Calhoon,
  Calvani, Canipe, Caputo, Cara, Carey, Case, Cesari, Cetorelli, Chance,
  Chandler, Chaney, Chapman, Charlot, Chayer, Cheezum, Chen, Chen, Cherinka,
  Chichester, Chilton, Chittiraibalan, Clampin, Clark, Clark, Clark,
  Claybrooks, Cleveland, Cohen, Cohen, Colón, Coleman, Colina, Comber, Comeau,
  Comer, Reis, Connolly, Conroy, Contos, Contreras, Cook, Cooper, Cooper,
  Correia, Correnti, Cossou, Costanza, Coulais, Cox, Coyle, Cracraft,
  Noriega-Crespo, Crew, Curtis, Cusveller, Maciel, Dailey, Daugeron, Davidson,
  Davies, Davis, Davis, Day, de~Chambure, de~Jong, De~Marchi, Dean, Decker,
  Delisa, Dell, Dellagatta, Dembinska, Demosthenes, Dencheva, Deneu, DePriest,
  Deschenes, Dethienne, Detre, Diaz, Dicken, DiFelice, Dillman, Disharoon, van
  Dishoeck, Dixon, Doggett, Dominguez, Donaldson, Doria-Warner, Santos, Doty,
  Douglas~Jr., Doyon, Dressler, Driggers, Driggers, Dunn, DuPrie, Dupuis,
  Durning, Dutta, Earl, Eccleston, Ecobichon, Egami, Ehrenwinkler, Eisenhamer,
  Eisenhower, Eisenstein, Hamel, Elie, Elliott, Elliott, Engesser, Espinoza,
  Etienne, Etxaluze, Evans, Fabreguettes, Falcolini, Falini, Fatig, Feeney,
  Feinberg, Fels, Ferdous, Ferguson, Ferrarese, Ferreira, Ferruit, Ferry,
  Filippazzo, Firre, Fix, Flagey, Flanagan, Fleming, Florian, Flynn, Foiadelli,
  Fontaine, Fontanella, Forshay, Fortner, Fox, Framarini, Francisco, Franck,
  Franx, Franz, Friedman, Friend, Frost, Fu, Fullerton, Gaillard, Galkin,
  Gallagher, Galyer, Marín, Gardner, Garland, Garrett, Gasman, Gáspár,
  Gastaud, Gaudreau, Gauthier, Geers, Geithner, Gennaro, Gerber, Gereau,
  Giampaoli, Giardino, Gibbons, Gilbert, Gilman, Girard, Giuliano, Gkountis,
  Glasse, Glassmire, Glauser, Glazer, Goldberg, Golimowski, Gonzaga, Gordon,
  Gordon, Goudfrooij, Gough, Graham, Grau, Green, Greene, Greene, Greenfield,
  Greenhouse, Greve, Greville, Grimaldi, Groe, Groebner, Grumm, Grundy, Güdel,
  Guillard, Guldalian, Gunn, Gurule, Gutman, Guy, Guyot, Hack, Haderlein,
  Hagan, Hagedorn, Hainline, Haley, Hami, Hamilton, Hammann, Hammel, Hanley,
  Hansen, Hardy, Harnisch, Harr, Harris, Hart, Hartig, Hasan, Hashim,
  Hashimoto, Haskins, Hawkins, Hayden, Hayden, Healy, Hecht, Heeg, Hejal, Helm,
  Hengemihle, Henning, Henry, Henry, Henshaw, Hernandez, Herrington, Heske,
  Hesman, Hickey, Hilbert, Hines, Hinz, Hirsch, Hitcho, Hodapp, Hodge, Hoffman,
  Holfeltz, Holler, Hoppa, Horner, Howard, Howard, Huber, Hunkeler, Hunter,
  Hunter, Hurd, Hurst, Hutchings, Hylan, Ignat, Illingworth, Irish, Isaacs~III,
  Jackson~Jr., Jaffe, Jahic, Jahromi, Jakobsen, James, James, James, Jamieson,
  Jandra, Jayawardhana, Jedrzejewski, Jeffers, Jensen, Joanne, Johns, Johnson,
  Johnson, Johnson, Johnson, Johnson, Johnson, Johnstone, Jollet, Jones, Jones,
  Jones, Jones, Jones, Jordan, Jordan, Jue, Jurkowski, Justis, Justtanont,
  Kaleida, Kalirai, Kalmanson, Kaltenegger, Kammerer, Kan, Kanarek, Kao,
  Karakla, Karl, Kassin, Kauffman, Kavanagh, Kelley, Kelly, Kendrew, Kennedy,
  Kenny, Keski-Kuha, Keyes, Khan, Kidwell, Kimble, King, King, Kinzel, Kirk,
  Kirkpatrick, Klaassen, Klingemann, Klintworth, Knapp, Knight, Knollenberg,
  Knutsen, Koehler, Koekemoer, Kofler, Kontson, Kovacs, Kozhurina-Platais,
  Krause, Kriss, Krist, Kristoffersen, Krogel, Krueger, Kulp, Kumari, Kwan,
  Kyprianou, Labador, Labiano, Lafrenière, Lagage, Laidler, Laine, Laird,
  Lajoie, Lallo, Lam, LaMassa, Lambros, Lampenfield, Lander, Langston, Larson,
  Larson, LaVerghetta, Law, Lawrence, Lee, Lee, Lee, Leisenring, Leveille,
  Levenson, Levi, Levine, Lewis, Lewis, Lewis, Libralato, Lidon, Liebrecht,
  Lightsey, Lilly, Lim, Lim, Ling, Link, Link, Lipinski, Liu, Lo, Lobmeyer,
  Logue, Long, Long, Long, Long, López-Caniego, Lotz, Love-Pruitt, Lubskiy,
  Luers, Luetgens, Luevano, Lui, Lund~III, Lundquist, Lunine, Lützgendorf,
  Lynch, MacDonald, MacDonald, Macias, Macklis, Maghami, Maharaja, Maiolino,
  Makrygiannis, Malla, Malumuth, Manjavacas, Marini, Marrione, Marston, Martel,
  Martin, Martin, Martinez, Maschmann, Masci, Masetti, Maszkiewicz, Matthews,
  Matuskey, McBrayer, McCarthy, McCaughrean, McClare, McClare, McCloskey,
  McClurg, McCoy, McElwain, McGregor, McGuffey, McKay, McKenzie, McLean,
  McMaster, McNeil, De~Meester, Mehalick, Meixner, Meléndez, Menzel, Menzel,
  Merz, Mesterharm, Meyer, Meyett, Meza, Midwinter, Milam, Miller, Miller,
  Miskey, Misselt, Mitchell, Mohan, Montoya, Moran, Morishita, Moro-Martín,
  Morrison, Morrison, Morse, Moschos, Moseley, Mosier, Mosner, Mountain,
  Muckenthaler, Mueller, Mueller, Muhiem, Mühlmann, Mullally, Mullen, Munger,
  Murphy, Murray, Muzerolle, Mycroft, Myers, Myers, Myers, Myers, Myrick,
  Nagle~IV, Nayak, Naylor, Neff, Nelan, Nella, Nguyen, Nguyen, Nickson,
  Nidhiry, Niedner, Nieto-Santisteban, Nikolov, Nishisaka, Noriega-Crespo,
  Nota, O'Mara, Oboryshko, O'Brien, Ochs, Offenberg, Ogle, Ohl, Olmsted,
  Osborne, O'Shaughnessy, Östlin, O'Sullivan, Otor, Ottens, Ouellette, Outlaw,
  Owens, Pacifici, Page, Paranilam, Park, Parrish, Paschal, Patapis, Patel,
  Patrick, Pattishall~Jr., Paul, Paul, Pauly, Pavlovsky, Peña-Guerrero,
  Pedder, Peek, Pelham, Penanen, Perriello, Perrin, Perrine, Perrygo, Peslier,
  Petach, Peterson, Pfarr, Pierson, Pietraszkiewicz, Pilchen, Pipher, Pirzkal,
  Pitman, Player, Plesha, Plitzke, Pohner, Poletis, Pollizzi, Polster, Pontius,
  Pontoppidan, Porges, Potter, Prescott, Proffitt, Pueyo, Neira, Radich, Rager,
  Rameau, Ramey, Alarcon, Rampini, Rapp, Rashford, Rauscher, Ravindranath,
  Rawle, Rawlings, Ray, Regan, Rehm, Rehm, Reid, Reis, Renk, Reoch, Ressler,
  Rest, Reynolds, Richon, Richon, Ridgaway, Riedel, Rieke, Rieke, Rifelli,
  Rigby, Riggs, Ringel, Ritchie, Rix, Robberto, Robinson, Robinson, Rock,
  Rodriguez, del Pino, Roellig, Rohrbach, Roman, Romelfanger, Romo~Jr.,
  Rosales, Rose, Roteliuk, Roth, Rothwell, Rouzaud, Rowe, Rowlands, Roy, Royer,
  Rui, Rumler, Rumpl, Russ, Ryan, Ryan, Saad, Sabata, Sabatino, Sabbi,
  Sabelhaus, Sabia, Sahu, Saif, Salvignol, Samara-Ratna, Samuelson, Sanders,
  Sappington, Sargent, Sauer, Savadkin, Sawicki, Schappell, Scheffer,
  Scheithauer, Scherer, Schiff, Schlawin, Schmeitzky, Schmitz, Schmude,
  Schneider, Schreiber, Schroeven-Deceuninck, Schultz, Schwab, Schwartz,
  Scoccimarro, Scott, Scott, Seaton, Seely, Seery, Seidleck, Sembach, Shanahan,
  Shaughnessy, Shaw, Shay, Sheehan, Sheth, Shih, Shivaei, Siegel, Sienkiewicz,
  Simmons, Simon, Sirianni, Sivaramakrishnan, Slade, Sloan, Slocum, Slowinski,
  Smith, Smith, Smith, Smith, Smith, Smith, Smolik, Soderblom, Sohn, Sokol,
  Sonneborn, Sontag, Sooy, Soummer, Southwood, Spain, Sparmo, Speer, Spencer,
  Sprofera, Stallcup, Stanley, Stansberry, Stark, Starr, Stassi, Steck,
  Steeley, Stephens, Stephenson, Stewart, Stiavelli, Stockman~Jr., Strada,
  Straughn, Streetman, Strickland, Strobele, Stuhlinger, Stys, Such, Sukhatme,
  Sullivan, Sullivan, Sumner, Sun, Sunnquist, Swade, Swam, Swenton, Swoish,
  Litten, Tamas, Tao, Taylor, Taylor, Plate, Van~Tea, Teague, Telfer, Temim,
  Texter, Thatte, Thompson, Thompson, Thomson, Thronson, Tierney, Tikkanen,
  Tinnin, Tippet, Todd, Tran, Trauger, Trejo, Truong, Tsukamoto, Tufail,
  Tumlinson, Tustain, Tyra, Ubeda, Underwood, Uzzo, Vaclavik, Valenduc,
  Valenti, Van~Campen, van~de Wetering, Van Der~Marel, van Haarlem,
  Vandenbussche, Vanterpool, Vernoy, Costas, Volk, Voorzaat, Voyton, Vydra,
  Waddy, Waelkens, Wahlgren, Walker~Jr., Wander, Warfield, Warner, Wasiak,
  Wasiak, Wehner, Weiler, Weilert, Weiss, Wells, Welty, Wheate, Wheeler, White,
  Whitehouse, Whiteleather, Whitman, Williams, Willmer, Willott, Willoughby,
  Wilson, Wilson, Wilson, Windhorst, Wislowski, Wolfe, Wolfe, Wolff, Wondel,
  Woo, Woods, Worden, Workman, Wright, Wu, Wu, Wun, Wymer, Yadetie, Yan, Yang,
  Yates, Yeager, Yerger, Young, Young, Yu, Yu, Zak, Zeidler, Zepp, Zhou,
  Zincke, Zonak, \& Zondag}]{gardner_james_2023}
Gardner, J.~P., Mather, J.~C., Abbott, R., {et~al.} 2023, The {James} {Webb}
  {Space} {Telescope} {Mission}, arXiv:2304.04869 [astro-ph]

\bibitem[{Gaspar {et~al.}(2023)Gaspar, Wolff, Rieke, Leisenring, Morrison, Su,
  Ward-Duong, Aguilar, Ygouf, Beichman, Llop-Sayson, \&
  Bryden}]{gaspar_spatially_2023}
Gaspar, A., Wolff, S.~G., Rieke, G.~H., {et~al.} 2023, Nature Astronomy,
  arXiv:2305.03789 [astro-ph]

\bibitem[{Hinkley {et~al.}(2022)Hinkley, Carter, Ray, Skemer, Biller, Choquet,
  Millar-Blanchaer, Sallum, Miles, Whiteford, Patapis, Perrin, Pueyo,
  Schneider, Stapelfeldt, Wang, Ward-Duong, Bowler, Boccaletti, Girard, Hines,
  Kalas, Kammerer, Kervella, Leisenring, Pantin, Zhou, Meyer, Liu, Bonnefoy,
  Currie, McElwain, Metchev, Wyatt, Absil, Adams, Barman, Baraffe, Bonavita,
  Booth, Bryan, Chauvin, Chen, Danielski, Furio, Factor, Fortney, Grady,
  Greenbaum, Henning, Janson, Kennedy, Kenworthy, Kraus, Kuzuhara, Lagage,
  Lagrange, Launhardt, Lazzoni, Lloyd, Marino, Marley, Martinez, Marois,
  Matthews, Matthews, Mawet, Phillips, Petrus, Quanz, Quirrenbach, Rameau,
  Rebollido, Rickman, Samland, Sargent, Schlieder, Sivaramakrishnan, Stone,
  Tamura, Tremblin, Uyama, Vasist, Vigan, Wagner, \& Ygouf}]{hinkley_jwst_2022}
Hinkley, S., Carter, A.~L., Ray, S., {et~al.} 2022, arXiv

\bibitem[{Hughes {et~al.}(2011)Hughes, Wilner, Andrews, Williams, Su,
  Murray-Clay, \& Qi}]{hughes_resolved_2011}
Hughes, A.~M., Wilner, D.~J., Andrews, S.~M., {et~al.} 2011, The Astrophysical
  Journal, 740, 38

\bibitem[{Ingraham {et~al.}(2014)Ingraham, Marley, Saumon, Marois, Macintosh,
  Barman, Bauman, Burrows, Chilcote, Rosa, Dillon, Doyon, Dunn, Erikson,
  Fitzgerald, Gavel, Goodsell, Graham, Hartung, Hibon, Kalas, Konopacky,
  Larkin, Maire, Marchis, McBride, Millar-Blanchaer, Morzinski, Norton,
  Oppenheimer, Palmer, Patience, Perrin, Poyneer, Pueyo, Rantakyro, Sadakuni,
  Saddlemyer, Savransky, Soummer, Sivaramakrishnan, Song, Thomas, Wallace,
  Wiktorowicz, \& Wolff}]{ingraham_gemini_2014}
Ingraham, P., Marley, M.~S., Saumon, D., {et~al.} 2014, The Astrophysical
  Journal Letters, 794, L15

\bibitem[{Konopacky {et~al.}(2013)Konopacky, Barman, Macintosh, \&
  Marois}]{konopacky_detection_2013}
Konopacky, Q.~M., Barman, T.~S., Macintosh, B.~A., \& Marois, C. 2013, Science,
  339, 1398

\bibitem[{Lacour {et~al.}(2019)Lacour, Nowak, Wang, Pfuhl, Eisenhauer, \&
  al.}]{lacour_first_2019}
Lacour, S., Nowak, M., Wang, J., {et~al.} 2019, Astronomy \& Astrophysics, 623,
  1

\bibitem[{{Lafreni{\`e}re} {et~al.}(2007){Lafreni{\`e}re}, {Marois}, {Doyon},
  {Nadeau}, \& {Artigau}}]{lafreniere_new_2007}
{Lafreni{\`e}re}, D., {Marois}, C., {Doyon}, R., {Nadeau}, D., \& {Artigau},
  {\'E}. 2007, \apj, 660, 770

\bibitem[{Lagrange {et~al.}(2023)Lagrange, Philipot, Rubini, Meunier, Kiefer,
  Kervella, Delorme, \& Beust}]{lagrange_radial_2023}
Lagrange, A.-M., Philipot, F., Rubini, P., {et~al.} 2023, On the radial
  distribution of giant exoplanets at {Solar} {System} scales, arXiv:2305.00047
  [astro-ph]

\bibitem[{{Lajoie} {et~al.}(2016){Lajoie}, {Soummer}, {Pueyo}, {Hines},
  {Nelan}, {Perrin}, {Clampin}, \& {Isaacs}}]{lajoie_small-grid_2016}
{Lajoie}, C.-P., {Soummer}, R., {Pueyo}, L., {et~al.} 2016, in Society of
  Photo-Optical Instrumentation Engineers (SPIE) Conference Series, Vol. 9904,
  Space Telescopes and Instrumentation 2016: Optical, Infrared, and Millimeter
  Wave, ed. H.~A. {MacEwen}, G.~G. {Fazio}, M.~{Lystrup}, N.~{Batalha},
  N.~{Siegler}, \& E.~C. {Tong}, 99045K

\bibitem[{Marois {et~al.}(2008)Marois, Macintosh, Barman, Zuckerman, Song,
  Patience, Lafreniere, \& Doyon}]{marois_direct_2008}
Marois, C., Macintosh, B., Barman, T., {et~al.} 2008, Science, 322, 1348

\bibitem[{Marois {et~al.}(2010)Marois, Zuckerman, Konopacky, Macintosh, \&
  Barman}]{marois_images_2010}
Marois, C., Zuckerman, B., Konopacky, Q.~M., Macintosh, B., \& Barman, T. 2010,
  Nature, 468, 1080

\bibitem[{Miles {et~al.}(2022)Miles, Biller, Patapis, Worthen, Rickman, Hoch,
  Skemer, Perrin, Chen, Mukherjee, Morley, Moran, Bonnefoy, Petrus, Carter,
  Choquet, Hinkley, Ward-Duong, Leisenring, Millar-Blanchaer, Pueyo, Ray,
  Stapelfeldt, Stone, Wang, Absil, Balmer, Boccaletti, Bonavita, Booth, Bowler,
  Chauvin, Christiaens, Currie, Danielski, Fortney, Girard, Greenbaum, Henning,
  Hines, Janson, Kalas, Kammerer, Kenworthy, Kervella, Lagage, Lew, Liu,
  Macintosh, Marino, Marley, Marois, Matthews, Matthews, Mawet, McElwain,
  Metchev, Meyer, Molliere, Pantin, Rebollido, Ren, Vasist, Wyatt, Zhou,
  Briesemeister, Bryan, Calissendorff, Catalloube, Cugno, De~Furio, Dupuy,
  Factor, Faherty, Fitzgerald, Franson, Gonzales, Hood, Howe, Kraus, Kuzuhara,
  Lawson, Lazzoni, Liu, Llop-Sayson, Lloyd, Martinez, Mazoyer, Quanz, Redai,
  Samland, Schlieder, Tamura, Tan, Uyama, Vigan, Vos, Wagner, Wolff, Ygouf,
  Zhang, \& Zhang}]{miles_jwst_2022}
Miles, B.~E., Biller, B.~A., Patapis, P., {et~al.} 2022, The {JWST} {Early}
  {Release} {Science} {Program} for {Direct} {Observations} of {Exoplanetary}
  {Systems} {II}: {A} 1 to 20 {Micron} {Spectrum} of the {Planetary}-{Mass}
  {Companion} {VHS} 1256-1257 b, arXiv:2209.00620 [astro-ph]

\bibitem[{Mollière {et~al.}(2020)Mollière, Stolker, Lacour, Otten, Shangguan,
  Charnay, Molyarova, Nowak, Henning, Marleau, Semenov, Van~Dishoeck,
  Eisenhauer, Garcia, Garcia~Lopez, Girard, Greenbaum, Hinkley, Kervella,
  Kreidberg, Maire, Nasedkin, Pueyo, Snellen, Vigan, Wang, De~Zeeuw, \&
  Zurlo}]{molliere_retrieving_2020}
Mollière, P., Stolker, T., Lacour, S., {et~al.} 2020, Astronomy \&
  Astrophysics, 640, A131

\bibitem[{Nielsen {et~al.}(2019)Nielsen, De~Rosa, Macintosh, Wang, Ruffio,
  Chiang, Marley, Saumon, Savransky, Mark~Ammons, Bailey, Barman, Blain,
  Bulger, Burrows, Chilcote, Cotten, Czekala, Doyon, Duchêne, Esposito,
  Fabrycky, Fitzgerald, Follette, Fortney, Gerard, Goodsell, Graham, Greenbaum,
  Hibon, Hinkley, Hirsch, Hom, Hung, Ilene~Dawson, Ingraham, Kalas, Konopacky,
  Larkin, Lee, Lin, Maire, Marchis, Marois, Metchev, Millar-Blanchaer,
  Morzinski, Oppenheimer, Palmer, Patience, Perrin, Poyneer, Pueyo, Rafikov,
  Rajan, Rameau, Rantakyrö, Ren, Schneider, Sivaramakrishnan, Song, Soummer,
  Tallis, Thomas, Ward-Duong, \& Wolff}]{nielsen_gemini_2019}
Nielsen, E.~L., De~Rosa, R.~J., Macintosh, B., {et~al.} 2019, The Astronomical
  Journal, 158, 13

\bibitem[{{Perrin} {et~al.}(2014){Perrin}, {Sivaramakrishnan}, {Lajoie},
  {Elliott}, {Pueyo}, {Ravindranath}, \& {Albert}}]{oschmann_updated_2014}
{Perrin}, M.~D., {Sivaramakrishnan}, A., {Lajoie}, C.-P., {et~al.} 2014, in
  Society of Photo-Optical Instrumentation Engineers (SPIE) Conference Series,
  Vol. 9143, Space Telescopes and Instrumentation 2014: Optical, Infrared, and
  Millimeter Wave, ed. J.~{Oschmann}, Jacobus~M., M.~{Clampin}, G.~G. {Fazio},
  \& H.~A. {MacEwen}, 91433X

\bibitem[{Petit Dit De La~Roche {et~al.}(2018)Petit Dit De La~Roche,
  Hoeijmakers, \& Snellen}]{petit_dit_de_la_roche_molecule_2018}
Petit Dit De La~Roche, D. J.~M., Hoeijmakers, H.~J., \& Snellen, I. A.~G. 2018,
  Astronomy \& Astrophysics, 616, A146

\bibitem[{Petit Dit De La~Roche {et~al.}(2020)Petit Dit De La~Roche,
  van den Ancker, Kissler-Patig, Ivanov, \&
  Fedele}]{petit_dit_de_la_roche_new_2020}
Petit Dit De La~Roche, D. J.~M., van den Ancker, M.~E., Kissler-Patig, M.,
  Ivanov, V.~D., \& Fedele, D. 2020, Monthly Notices of the Royal Astronomical
  Society, 491, 1795

\bibitem[{Petrus {et~al.}(2023)Petrus, Chauvin, Bonnefoy, Tremblin, Charnay,
  Delorme, Marleau, Bayo, Manjavacas, Lagrange, Mollière, Palma-Bifani,
  Biller, Jenkins, Goyal, \& Hoch}]{petrus_x-shyne_2023}
Petrus, S., Chauvin, G., Bonnefoy, M., {et~al.} 2023, Astronomy \&
  Astrophysics, 670, L9

\bibitem[{Phillips {et~al.}(2020)Phillips, Tremblin, Baraffe, Chabrier, Allard,
  Spiegelman, Goyal, Drummond, \& Hébrard}]{phillips_new_2020}
Phillips, M.~W., Tremblin, P., Baraffe, I., {et~al.} 2020, Astronomy \&
  Astrophysics, 637, A38

\bibitem[{Polletta {et~al.}(2007)Polletta, Tajer, Maraschi, Trinchieri,
  Lonsdale, Chiappetti, Andreon, Pierre, Le~Fevre, Zamorani, Maccagni, Garcet,
  Surdej, Franceschini, Alloin, Shupe, Surace, Fang, Rowan‐Robinson, Smith,
  \& Tresse}]{polletta_spectral_2007}
Polletta, M., Tajer, M., Maraschi, L., {et~al.} 2007, The Astrophysical
  Journal, 663, 81

\bibitem[{Pueyo(2016)}]{pueyo_detection_2016}
Pueyo, L. 2016, The Astrophysical Journal, 824, 117

\bibitem[{Read {et~al.}(2018)Read, Wyatt, Marino, \&
  Kennedy}]{read_shaping_2018}
Read, M.~J., Wyatt, M.~C., Marino, S., \& Kennedy, G.~M. 2018, Monthly Notices
  of the Royal Astronomical Society, 475, 4953

\bibitem[{Rigby {et~al.}(2022)Rigby, Perrin, McElwain, Kimble, Friedman, Lallo,
  Doyon, Ferruit, Glasse, Rieke, Rieke, Wright, Willott, Colon, Milam, Neff,
  Stark, Valenti, Abell, Abney, Abul-Huda, Acton, Adams, Adler, Aguilar, Ahmed,
  Albert, Alberts, Aldridge, Allen, Altenburg, Álvarez Márquez, de~Oliveira,
  Andersen, Anderson, Argyriou, Armstrong, Arribas, Artigau, Arvai, Atkinson,
  Bacon, Bair, Banks, Barrientes, Barringer, Bartosik, Bast, Beatty, Bechtold,
  Beck, Bergeron, Bergkoetter, Bhatawdekar, Birkmann, Blazek, Blome,
  Boccaletti, Böker, Boia, Bonaventura, Bond, Bosley, Boucarut, Bourque,
  Bouwman, Bower, Bowers, Boyer, Bradley, Brady, Braun, Breda, Bresnahan,
  Bright, Britt, Bromenschenkel, Brooks, Brooks, Brown, Brown, Brown, Bunker,
  Burger, Bushouse, Cale, Cameron, Cameron, Canipe, Caputo, Cara, Carey,
  Carniani, Carrasquilla, Carruthers, Case, Catherine, Chance, Chapman,
  Charlot, Charlow, Chayer, Cherinka, Chichester, Chilton, Chonis, Clampin,
  Clark, Clark, Coleman, Comber, Comeau, Connolly, Cooper, Cooper, Coppock,
  Correnti, Cossou, Coulais, Coyle, Cracraft, Curti, Cuturic, Davis, Davis,
  Dean, DeLisa, DePasquale, Deschenes, Detre, Diaz, Dicken, DiFelice, Dillman,
  Dixon, Doggett, Donaldson, Douglas, DuPrie, Dupuis, Durning, Eck, Edeani,
  Egami, Ehrenwinkler, Eisenhamer, Eisenhower, Elie, Elliott, Elliott, Ellis,
  Engesser, Espinoza, Etienne, Falini, Feeney, Ferry, Filippazzo, Fincham, Fix,
  Florian, Flynn, Fontanella, Ford, Forshay, Fox, Franz, Fu, Fullerton, Galkin,
  Galyer, Marin, Gardner, Gardner, Garland, Garrett, Gasman, Gaspar, Gaudreau,
  Gauthier, Geers, Geithner, Gennaro, Giardino, Girard, Giuliano, Glassmire,
  Glauser, Glazer, Godfrey, Golimowski, Gollnitz, Gong, Gonzaga, Gordon,
  Goudfrooij, Greene, Greenhouse, Grimaldi, Groebner, Guillard, Gutman, Ha,
  Haderlein, Hagedorn, Hainline, Haley, Hami, Hamilton, Hammel, Hansen,
  Harkins, Harr, Hart, Hartig, Hashimoto, Haskins, Hathaway, Havey, Hayden,
  Hecht, Heller-Boyer, Henriques, Henry, Hermann, Hernandez, Hesman, Hilbert,
  Hines, Hoffman, Holfeltz, Holler, Hoppa, Hott, Howard, Howard, Hunter,
  Hunter, Hurst, Husemann, Hustak, Ignat, Illingworth, Irish, Jackson, Jahromi,
  Jakobsen, James, Januszewski, Jenkins, Jirdeh, Johnson, Johnson, Jones,
  Jones, Jones, Jordan, Jordan, Jurczyk, Jurling, Kalmanson, Kammerer, Kang,
  Kao, Karakla, Kavanagh, Kendrew, Kennedy, Kenny, Keski-kuha, Keyes, Kidwell,
  Kirk, Kirkpatrick, Kirshenblat, Klaassen, Knapp, Knight, Koehler, Koekemoer,
  Kovacs, Kulp, Kumari, Kyprianou, Massa, Labador, Labiano, Lagage, Lajoie,
  Lam, Lamb, Lambros, Lampenfield, Langston, Larson, Law, Lee, Leisenring,
  Lepo, Leveille, Levenson, Levine, Levy, Lewis, Libralato, Lightsey, Link,
  Liu, Lo, Lockwood, Long, Long, Loomis, Lopez-Caniego, Alvarez, \&
  Love-Pruitt}]{rigby_science_2022}
Rigby, J., Perrin, M., McElwain, M., {et~al.} 2022, arXiv

\bibitem[{Ruffio {et~al.}(2021)Ruffio, Konopacky, Barman, Macintosh, Hoch,
  Rosa, Wang, Czekala, \& Marois}]{ruffio_deep_2021}
Ruffio, J.-B., Konopacky, Q.~M., Barman, T., {et~al.} 2021, The Astronomical
  Journal, 162, 290

\bibitem[{Skaf {et~al.}(2022)Skaf, Guyon, Gendron, Ahn, Bertrou-Cantou,
  Boccaletti, Cranney, Currie, Deo, Edwards, Ferreira, Gratadour, Lozi, Norris,
  Sevin, Vidal, \& Vievard}]{skaf_-sky_2022}
Skaf, N., Guyon, O., Gendron, E., {et~al.} 2022, Astronomy \& Astrophysics,
  659, A170

\bibitem[{Skemer {et~al.}(2014)Skemer, Marley, Hinz, Morzinski, Skrutskie,
  Leisenring, Close, Saumon, Bailey, Briguglio, Defrere, Esposito, Follette,
  Hill, Males, Puglisi, Rodigas, \& Xompero}]{skemer_directly_2014}
Skemer, A.~J., Marley, M.~S., Hinz, P.~M., {et~al.} 2014, The Astrophysical
  Journal, 792, 17

\bibitem[{{Soummer} {et~al.}(2014){Soummer}, {Lajoie}, {Pueyo}, {Hines},
  {Isaacs}, {Nelan}, {Clampin}, \& {Perrin}}]{soummer_small-grid_2014}
{Soummer}, R., {Lajoie}, C.-P., {Pueyo}, L., {et~al.} 2014, in Society of
  Photo-Optical Instrumentation Engineers (SPIE) Conference Series, Vol. 9143,
  Space Telescopes and Instrumentation 2014: Optical, Infrared, and Millimeter
  Wave, ed. J.~{Oschmann}, Jacobus~M., M.~{Clampin}, G.~G. {Fazio}, \& H.~A.
  {MacEwen}, 91433V

\bibitem[{Soummer {et~al.}(2012)Soummer, Pueyo, \&
  Larkin}]{soummer_detection_2012}
Soummer, R., Pueyo, L., \& Larkin, J. 2012, The Astrophysical Journal Letters,
  755, L28

\bibitem[{Su {et~al.}(2009)Su, Rieke, Stapelfeldt, Malhotra, Bryden, Smith,
  Misselt, Moro-Martín, \& Williams}]{su_debris_2009}
Su, K. Y.~L., Rieke, G.~H., Stapelfeldt, K.~R., {et~al.} 2009, The
  Astrophysical Journal, 705, 314

\bibitem[{Tremblin {et~al.}(2016)Tremblin, Amundsen, Chabrier, Baraffe,
  Drummond, Hinkley, Mourier, \& Venot}]{tremblin_cloudless_2016}
Tremblin, P., Amundsen, D.~S., Chabrier, G., {et~al.} 2016, The Astrophysical
  Journal, 817, L19

\bibitem[{Tremblin {et~al.}(2017)Tremblin, Chabrier, Baraffe, Liu, Magnier,
  Lagage, Oliveira, Burgasser, Amundsen, \& Drummond}]{tremblin_cloudless_2017}
Tremblin, P., Chabrier, G., Baraffe, I., {et~al.} 2017, The Astrophysical
  Journal, 850, 46

\bibitem[{Vigan {et~al.}(2020)Vigan, Fontanive, Meyer, Biller, Bonavita, Feldt,
  Desidera, Marleau, Emsenhuber, Galicher, Rice, Forgan, Mordasini, Gratton,
  Coroller, Maire, Cantalloube, Chauvin, Cheetham, Hagelberg, Lagrange,
  Langlois, Bonnefoy, Beuzit, Boccaletti, D’Orazi, Delorme, Dominik, Henning,
  Janson, Lagadec, Lazzoni, Ligi, Menard, Mesa, Messina, Moutou, Müller,
  Perrot, Samland, Schmid, Schmidt, Sissa, Turatto, Udry, Zurlo, Abe, Antichi,
  Asensio-Torres, Baruffolo, Baudoz, Baudrand, Bazzon, Blanchard, Bohn,
  Sevilla, Carbillet, Carle, Cascone, Charton, Claudi, Costille, Caprio,
  Delboulbé, Dohlen, Engler, Fantinel, Feautrier, Fusco, Gigan, Girard, Giro,
  Gisler, Gluck, Gry, Hubin, Hugot, Jaquet, Kasper, Mignant, Llored, Madec,
  Magnard, Martinez, Maurel, Möller-Nilsson, Mouillet, Moulin, Origné,
  Pavlov, Perret, Petit, Pragt, Puget, Rabou, Ramos, Rickman, Rigal, Rochat,
  Roelfsema, Rousset, Roux, Salasnich, Sauvage, Sevin, Soenke, Stadler, Suarez,
  Wahhaj, Weber, \& Wildi}]{vigan_sphere_2020}
Vigan, A., Fontanive, C., Meyer, M., {et~al.} 2020, Astronomy \& Astrophysics,
  651, A72

\bibitem[{Wagner {et~al.}(2021)Wagner, Boehle, Pathak, Kasper, Arsenault,
  Jakob, Käufl, Leveratto, Maire, Pantin, Siebenmorgen, Zins, Absil, Ageorges,
  Apai, Carlotti, Choquet, Delacroix, Dohlen, Duhoux, Forsberg, Fuenteseca,
  Gutruf, Guyon, Huby, Kampf, Karlsson, Kervella, Kirchbauer, Klupar, Kolb,
  Mawet, N’Diaye, Xivry, Quanz, Reutlinger, Ruane, Riquelme, Soenke, Sterzik,
  Vigan, \& Zeeuw}]{wagner_imaging_2021}
Wagner, K., Boehle, A., Pathak, P., {et~al.} 2021, Nature Communications, 12,
  922

\bibitem[{Wang {et~al.}(2022)Wang, Gao, Chilcote, Lozi, Guyon, Marois, De~Rosa,
  Sahoo, Groff, Vievard, Jovanovic, Greenbaum, \&
  Macintosh}]{wang_atmospheric_2022}
Wang, J.~J., Gao, P., Chilcote, J., {et~al.} 2022, The Astronomical Journal,
  164, 143

\bibitem[{Wang {et~al.}(2021)Wang, Kulikauskas, \&
  Blunt}]{wang_whereistheplanet_2021}
Wang, J.~J., Kulikauskas, M., \& Blunt, S. 2021, Astrophysics Source Code
  Library, ascl:2101.003, aDS Bibcode: 2021ascl.soft01003W

\bibitem[{Wilner {et~al.}(2018)Wilner, MacGregor, Andrews, Hughes, Matthews, \&
  Su}]{wilner_resolved_2018}
Wilner, D.~J., MacGregor, M.~A., Andrews, S.~M., {et~al.} 2018, The
  Astrophysical Journal, 855, 56

\bibitem[{Wright {et~al.}(2015)Wright, Wright, Goodson, Rieke, Aitink-Kroes,
  Amiaux, Aricha-Yanguas, Azzollini, Banks, Barrado-Navascues,
  Belenguer-Davila, Bloemmart, Bouchet, Brandl, Colina, Detre, Diaz-Catala,
  Eccleston, Friedman, García-Marín, Güdel, Glasse, Glauser, Greene,
  Groezinger, Grundy, Hastings, Henning, Hofferbert, Hunter, Jessen,
  Justtanont, Karnik, Khorrami, Krause, Labiano, Lagage, Langer, Lemke, Lim,
  Lorenzo-Alvarez, Mazy, McGowan, Meixner, Morris, Morrison, Müller, rgaard
  Nielson, Olofsson, O'Sullivan, Pel, Penanen, Petach, Pye, Ray, Renotte,
  Renouf, Ressler, Samara-Ratna, Scheithauer, Schneider, Shaughnessy,
  Stevenson, Sukhatme, Swinyard, Sykes, Thatcher, Tikkanen, Dishoeck, Waelkens,
  Walker, Wells, \& Zhender}]{wright_mid-infrared_2015}
Wright, G.~S., Wright, D., Goodson, G.~B., {et~al.} 2015, Publications of the
  Astronomical Society of the Pacific, 127, 595

\bibitem[{Öberg {et~al.}(2011)Öberg, Murray-Clay, \&
  Bergin}]{oberg_effects_2011}
Öberg, K.~I., Murray-Clay, R., \& Bergin, E.~A. 2011, The Astrophysical
  Journal, 743, L16

\end{thebibliography}

\end{document}